\documentclass[journal=jacsat, superscriptaddress]{achemso}

\usepackage{graphicx}
\usepackage{amsmath}
\usepackage{amssymb}
\SectionNumbersOff
\setkeys{acs}{articletitle = true}

\usepackage{sectsty}
\usepackage{enumitem}
\usepackage{color}
\SectionsOn

\usepackage[normalem]{ulem}

\newif\ifHighlitedChanges
\def\ifHighlitedChanges{\iftrue}
\ifHighlitedChanges
  \def\EDITS#1{{\color{black}#1}}
  \def\STRIKE#1{{\color{black}\sout{#1}}}
\else
  \def\EDITS#1{#1}
  \def\STRIKE#1{\relax}
\fi

\newif\ifJCCHighlitedChanges
\def\ifJCCHighlitedChanges{\iftrue}
\ifJCCHighlitedChanges
  \def\EDITSJCC#1{{\color{black}#1}}
  \def\STRIKEJCC#1{{\color{black}\sout{#1}}}
\else
  \def\EDITSJCC#1{#1}
  \def\STRIKEJCC#1{\relax}
\fi

\newif\ifRKHighlitedChanges
\def\ifRkHighlitedChanges{\iftrue}
\ifRKHighlitedChanges
  
  \def\STRIKERK#1{{\color{black}\sout{#1}}}
\else
  
  \def\STRIKERK#1{\relax}
\fi

\title{Tunable Assembly of Gold Nanorods in Polymer Solutions to Generate Controlled Nanostructured Materials}

\author{Ryan Poling-Skutvik}

\affiliation{Department of Chemical and Biomolecular Engineering, University of Houston, Houston, TX 77204-4004}

\author{Jonghun Lee}
\affiliation{Advanced Photon Source, Argonne National Laboratory, Argonne, IL 60439}

\author{Suresh Narayanan}
\affiliation{Advanced Photon Source, Argonne National Laboratory, Argonne, IL 60439}

\author{Ramanan Krishnamoorti}
\email{ramanan@uh.edu}

\affiliation{Department of Chemical and Biomolecular Engineering, University of Houston, Houston, TX 77204-4004}

\author{Jacinta C. Conrad}
\email{jcconrad@uh.edu}

\affiliation{Department of Chemical and Biomolecular Engineering, University of Houston, Houston, TX 77204-4004}

\date{\today}

\begin{document}
\noindent\EDITSJCC{Keywords: depletion interactions}; \EDITSJCC{fractal structures}; \EDITSJCC{protein limit}; \EDITSJCC{small-angle x-ray scattering}; \EDITSJCC{optical absorption spectroscopy}


\begin{abstract}

Gold nanorods grafted with short chain polymers are assembled into controlled open structures using polymer-induced depletion interactions and structurally characterized using small angle x-ray scattering. When the nanorod diameter is smaller than the radius of gyration of the depletant polymer, the depletion interaction depends solely on the correlation length of the polymer solution and not directly on the polymer molecular weight. As the polymer concentration increases, the stronger depletion interactions increasingly compress the  grafted chains and push the gold nanorods closer together. By contrast, other structural characteristics such as the number of nearest neighbors and fractal dimension exhibit a non-monotonic dependence on polymer concentration. These parameters are maximal at intermediate concentrations, which are attributed to a crossover from reaction-limited to diffusion-limited aggregation. The control over structural properties of anisotropic nanoscale building blocks demonstrated here will be beneficial to designing and producing materials \emph{in situ} with specific direction-dependent nanoscale properties and provides a crucial route for advances in additive manufacturing.

\end{abstract}

\section{Introduction}

Gold nanoparticles exhibit unique optical and electronic properties, advantageous for applications ranging from drug delivery and theranostics\cite{Alkilany2010,Choi2011} to sensing\cite{Saha2012} to electronics\cite{Romo-Herrera2011} to catalysis.\cite{Thompson2007} Suspending the nanoparticles in solutions maximizes their accessible surface area to take advantage of these novel properties. Maximizing surface area requires that the nanoparticles remain stable and dispersed as individual particles, often in the presence of environmental factors that alter stability such as pH,\cite{Orendorff2005,Nam2009} ionic strength,\cite{Boles2016} and macromolecular depletants\cite{Tam2010,Murthy2013}. Hence, in many applications the nanoparticle surface is functionalized with surfactants, charged compounds, or macromolecules to induce repulsions between the nanoparticles or favorable chemical interactions with the surrounding solution. In other settings, however, multiparticle assemblies of gold nanoparticles exhibit distinctive and desirable functional properties, such as modified cellular uptake\cite{Nam2009,Albanese2011} or binding and detecting biomolecules.\cite{Schwartzberg2004} \EDITSJCC{As a second example, percolating nanostructures can improve the thermal or electrical conductivity and optical or mechanical properties of composite materials.\cite{Paul2008,Kumar2010}} Tuning the desired functional properties requires control over the assembly of the nanoparticles. Spherical nanoparticles are easily assembled into amorphous aggregates,\cite{Dimon1986,Lin1989,Boal2000} but the production of one and two-dimensional structures requires patchy functionalization.\cite{Akcora2009,Kumar2013} High-aspect-ratio nanorods such as carbon nanotubes can pack into triangular or hexagonal structures,\cite{Frank1998,Baranov2010,Park2010} and form liquid crystals at modest particle volume fractions.\cite{Song2003,Rai2006}  Nanoparticles with greater geometric complexity can form a variety of complex superlattice structures due to the directionality of attractive interactions.\cite{Zanella2011,Young2013} Open fractal structures featuring a hierarchy of length scales are crucial to applications such as additive manufacturing\cite{Smay2002} and have been observed for polymer nanocomposites prepared with carbon nanotubes but only at high nanoparticle concentrations.\cite{Chatterjee2008} The production of controlled and tunable 3-D structures of anisotropic nanoparticles at low concentrations, however, continues to pose a challenge. 

One route to assemble nanoparticles is to induce well-controlled depletion interactions from macromolecules in solution. Polymers added to solution, as one example, have well-defined characteristic length scales (such as the radius of gyration $R_\mathrm{g}$ and the correlation length $\xi$ representing the average distance between neighboring chains). Polymer-induced depletion interactions are well studied for micron-sized colloidal particles,\cite{Lekkerkerker2011} where the polymer chains are much smaller than the colloids, but additional factors alter interactions between nanoscale particles in crowded macromolecular solutions. The simplest theory for depletion interactions, due to Asakura and Oosawa,\cite{Asakura1958} approximates the polymer coils as non-interacting hard spheres so that the depletant concentration is negligible at the particle surface and discontinuously increases to the bulk concentration at a distance equal to the depletant radius away from the particle surface. Under this assumption, the interaction potential between particles is proportional to the osmotic pressure of the solution and the volume excluded to the polymer. The flexibility of the polymer chains, however, smooths out the concentration profile of the depletant near the particle surface, and can thus modulate the strength and range of polymer-induced forces.\cite{Kleshchanok2008} To a first order approximation, the depletion strength becomes inversely proportional to $\xi$ in semidilute polymer solutions.\cite{DeGennes1979} Furthermore, enthalpic interactions between the polymeric depletants and particles affect the interparticle forces, leading to stronger attraction for unfavorable enthalpic interactions and weaker attraction or even repulsion for favorable particle-polymer interactions.\cite{Semenov2015} Because nanoparticles are much smaller than colloids, their characteristic length scales are comparable in magnitude to those of the depleting polymer. In this ``protein limit,''\cite{Mutch2007} depletion interactions may exhibit further deviations from the simple Asakura-Oosawa picture,\cite{Kulkarni1999,Vivares2002} instead agreeing with predictions from the polymer reference-interaction site model.\cite{Chatterjee1998} These deviations modify the phase behavior of suspensions of nearly-spherical nanoparticles,\cite{Wormutht1991,Clegg1994,Piculell1996,Tuinier2000,Ramakrishnan2002a,Ramakrishnan2002b,Vliegenthart2003,Hennequin2005} but the effect of depletion interactions in the ``protein limit'' on the assembly of anisotropic particles remains incompletely understood.

Here, we use a well-characterized model system of polymer-grafted gold nanorods (AuNRs) in polymer solutions to investigate the effects of macromolecular crowding on the structure of AuNR aggregates. AuNRs synthesized through a seed-mediated growth method are grafted with short poly(ethylene oxide) (PEO) chains and suspended in semidilute solutions of PEO of varying molecular weight. The optical spectra of these suspensions, arising from localized surface plasmon resonances, exhibit marked shifts as a function of solution PEO size and concentration, consistent with changes in the dispersion of the AuNRs. Small-angle x-ray scattering (SAXS) experiments reveal that the AuNRs aggregate at sufficiently high concentrations of  polymer. The characteristic distance between the AuNRs decreases as the concentration of polymer in solution is increased, independent of the molecular weight of the depletant. These changes in structure are consistent with an increase in depletion strength that compresses the grafted brushes, controlled only by the correlation length $\xi$ of polymers in solution. Both the fractal dimension (determined from the low-wavevector slope of the structure factor) and the number of neighbors (assessed semi-quantitatively from the ratio of the nearest neighbor peak height) depend non-monotonically on polymer concentration and hence polymer correlation length $\xi$, suggesting that the aggregate structure is the result of opposing kinetic processes. The dynamics of nanoparticles in polymer solutions decouples from the bulk viscoelasticity of the polymer solutions and is controlled by the nanoparticle diameter $d_\mathrm{NP}$ and $\xi$.\cite{Cai2011,Poling-Skutvik2015} The dependence of both attraction strength and particle mobility on $\xi$ collapses the structural properties onto a single master curve. Thus the structure is determined by a competition between the attraction strength and particle transport rate, which respectively increase and decrease as $\xi$ decreases, resulting in a transition from reaction-limited to diffusion-limited aggregation with increasing polymer concentration. 

\section{Materials and Methods}

\subsection{\small Gold nanorod synthesis}

Gold nanorods (AuNRs) are synthesized using a seed-mediated growth method.\cite{Ye2012} Briefly, a seed solution is prepared by mixing 5 mL of 0.5 mM chloroauric acid (HAuCl$_4$, Sigma-Aldrich) with 5 mL of 0.2 M cetyl trimethylammonium bromide (CTAB, Alfa Aesar) to which 0.6 mL of fresh 0.01 M sodium borohydride (NaBH$_4$, Sigma-Aldrich) solution is added to initiate seed growth. To synthesize the AuNRs, a mixture of 4.5 g of CTAB, 0.55 g of 5-bromosalicylic acid (Alfa Aesar), and 125 mL of water is heated to 70 $^\circ$C to dissolve the 5-bromosalicylic acid and then cooled to 30 $^\circ$C for the rest of the reaction. Silver nitrate (6 mL \EDITS{at a concentration of 4 mM}, Sigma-Aldrich) is added and left unagitated for 15 minutes. Next, 125 mL of 1 mM HAuCl$_4$ is added and stirred for 15 minutes. Finally, 1 mL of \EDITS{64 mM} L-ascorbic acid (Sigma-Aldrich) and 0.4 mL of the seed solution are added, and the reaction mixture is briefly stirred for 30 seconds before reacting undisturbed for 12 hours. The resulting AuNRs are purified by centrifugation, decanting supernatant, and redispersing in fresh water twice. To ensure a neutral interaction between the AuNRs and the poly(ethylene oxide) (PEO) in solution, the purified AuNRs are reacted with 4 g of thiol-PEO (weight-averaged molecular weight $M_\mathrm{W} = 2$ kDa, Nanocs), which covalently bonds to the gold surface, for over 24 hours at room temperature. The functionalized AuNRS are repurified using the same method. The dimensions of the functionalized AuNRs are $58 \pm 5$ nm in length and $17 \pm 3$ nm in diameter (aspect ratio $L/d_\mathrm{NP} = 3.4 \pm 0.7$), as measured using transmission electron microscopy (TEM). \EDITS{Due to the low $M_\mathrm{W}$ of the thiol-PEO and the high mass of the AuNRs, the thiol-PEO grafting density cannot be determined through traditional methods such as thermal gravimetric analysis. Previous studies,\cite{Harder1998,Rahme2013} however, \EDITSJCC{reported} grafting densities for short-chain thiol-PEO \EDITSJCC{of} $\approx 3-4$ chains nm$^{-2}$.  }

\subsection{\small Rheology}

Initial polymer stock solutions are prepared by dissolving the appropriate amount of PEO ($M_\mathrm{W} = 35$, 100, 200, 400, or 1000 kDa) in water (Milli-Q, Millipore) and homogenizing on a roll mixer for 2 days. The stock solution is then diluted with additional water and further homogenized for 24 hours to produce solutions with the desired polymer concentration. Rheology experiments are performed on a Discovery Hybrid rheometer (DHR-2, TA Instruments) at constant stress in the steady-shear configuration using a Couette geometry with a bob length of 42 mm and diameter of 28 mm. The steady-shear viscosities are measured as a function of shear rate to ensure that the Newtonian zero-shear plateau is reached. 

\subsection{\small Small-angle x-ray scattering }

Polymer solutions are prepared as described for rheology experiments. A small amount of a concentrated suspension of AuNRs (volume fraction $\phi \approx 10^{-3}$) is added to the polymer solution and mixed with a vortex mixer for solutions with polymer concentration $c < 10c^*$ or stirred manually until homogeneous for higher concentrations to produce a suspension with an AuNR concentration of $\phi \approx 10^{-5}$. This AuNR concentration is high enough to generate significant scattering intensity but low enough to mitigate interparticle interactions or multiple scattering effects. Prepared solutions are pipetted into 1-mm o.d.\ quartz capillaries, which are briefly centrifuged to drive the solution to the bottom and then sealed with wax to prevent evaporation. We collect small angle x-ray scattering (SAXS) data over a wavevector range 0.0019 \AA$^{-1}$ $< Q <$ 0.08 {\AA}$^{-1}$, corresponding to length scales ranging from approximately 10 nm to 300 nm, at the 8-ID-I beamline at the Advanced Photon Source, Argonne National Laboratory. \EDITS{The 2-D scattering intensities for all solutions are azimuthally uniform and show no signature of alignment or preferred orientation. Thus, the 2-D scattering intensity is azimuthally averaged into a 1-D scattering intensity as a function of scattering wavevector $Q$.}

\subsection{\small UV spectroscopy}

The AuNR suspensions used for XPCS and SAXS measurements are diluted further with additional polymer solution for a final AuNR concentration $\phi \approx 10^{-7}$. Solutions with higher AuNR concentrations are too optically dense, precluding measurements of their optical spectra. \EDITS{Optical absorption spectra} are collected on a Jasco V-570 spectrophotometer using the neat polymer solution as a reference.

\section{Results}

\subsection{\small Characterization of polymer solutions}

\begin{figure}[ht!]
\includegraphics[width = 2.9 in]{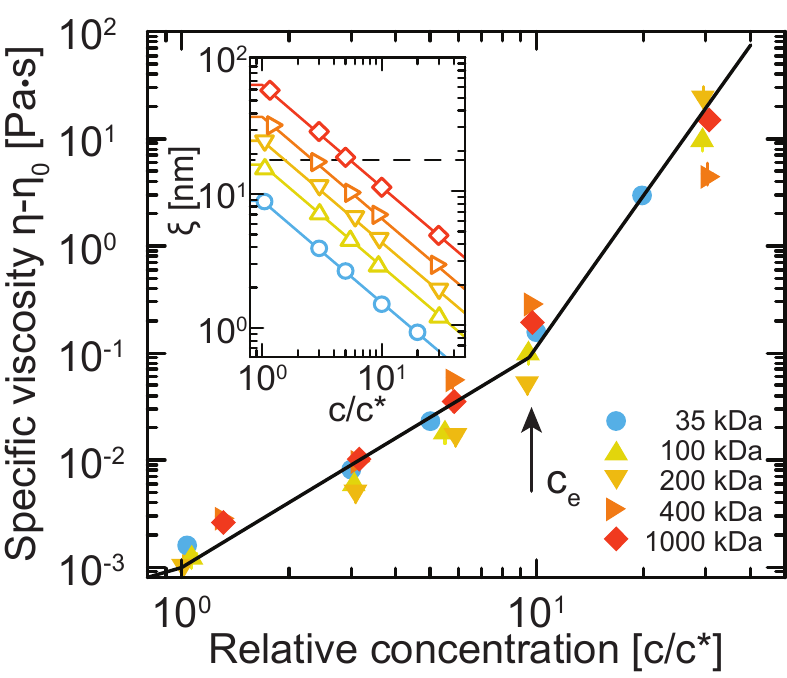}
\caption{\label{fig:Rheology} Bulk specific viscosity $\eta - \eta_0$ as a function of relative polymer concentration $c/c^*$ for various PEO molecular weights. Solid lines are scaling predictions\cite{Rubinstein2003} with an entanglement concentration $c_\mathrm{e} \approx 9.5 c^*$. \textit{Inset}: Predicted correlation length $\xi$ for the bulk rheology samples. Dashed line indicates the diameter $d_\mathrm{NP}$ of AuNRs.}
\end{figure}

The viscoelastic properties of polymer solutions are dictated by two characteristic length scales: the polymer radius of gyration $R_\mathrm{g}$ and the correlation length $\xi$ between chains. In the dilute limit, polymer chains exist as individual Gaussian chains with $R_\mathrm{g,0} = \left[ M_\mathrm{w}/(4/3\pi N_\mathrm{av}c^*) \right]^{1/3}$, where $M_\mathrm{w}$ is the molecular weight and $N_\mathrm{av}$ is Avogadro's number. \EDITS{$R_\mathrm{g,0}$ are calculated to be 8.5, 16, 24, 36, and 62 nm for $M_\mathrm{W} = 35$, 100, 200, 400, and 1000 kDa PEO, respectively.} Above the overlap concentration $c^*$, the chains interact so that a second length scale $\xi$ develops and scales with concentration according to $\xi = R_\mathrm{g,0}(c/c^*)^{-\nu/(3\nu-1)}$, where $\nu = 0.59$ is the excluded volume exponent for good solvent conditions \EDITS{(Table S1, Supporting Information)}. Scaling theory\cite{Rubinstein2003} predicts that the bulk zero-shear viscosity of the polymer solutions should collapse onto a single master curve as a function of relative polymer concentration $c/c^*$, as confirmed for the PEO solutions using bulk rheology (Fig.\ \ref{fig:Rheology}). The viscosity of the solutions scales as $\eta-\eta_0 \sim (c/c^*)^2$ until the entanglement concentration $c_e$, beyond which $\eta-\eta_0 \sim (c/c^*)^{14/3}$. Although the bulk viscosities collapse onto a master curve as a function of relative polymer concentrations, the individual length scales do not (inset to Fig.\ \ref{fig:Rheology}). Thus, these polymer solutions have similar bulk viscoelastic properties but substantially different nanoscale structures.

\subsection{Optical properties of AuNR suspensions}

\begin{figure*}[b!]
\includegraphics[width = 6.5 in]{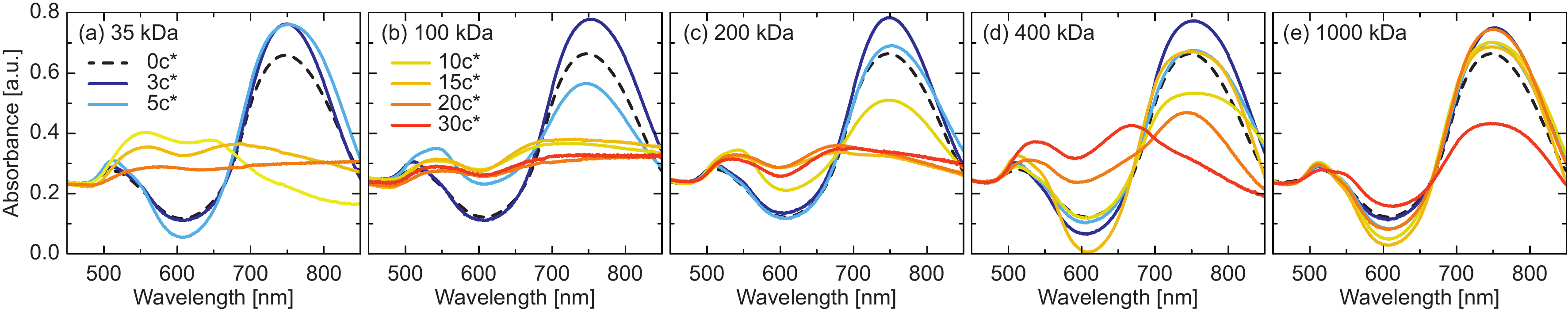}
\caption{\label{fig:uvvis} \EDITS{Optical absorption} spectra for suspensions of AuNRs in solutions of (a) 35 kDa, (b) 100 kDa, (c) 200 kDa, (d) 400 kDa, and (e) 1000 kDa PEO at various concentrations. Curves are shifted vertically to overlap at 450 nm.}
\end{figure*}

The optical properties of AuNRs depend on the shape and dispersion of the gold nanorods. Using optical absorption spectroscopy, we assess the dispersion of the AuNRs in the polymer solutions. Whereas AuNRs used in many previous studies are stabilized with electrostatic charges\cite{Orendorff2005,Park2010b,Choi2011}, the AuNRs used in this study are covalently functionalized with short 2 kDa PEO chains. With this functionalization, the AuNRs are stable and well-dispersed in water, and the absorption spectrum exhibits two local maxima at wavelengths of 510 and 745 nm, corresponding to the transverse and longitudinal surface plasmon resonances, respectively (Fig.\ \ref{fig:uvvis}). Thus, the surface functionalization appears to be uniform across the particle surface, ensuring neutral interactions between AuNRs and the dissolved PEO chains. For AuNRs dispersed in polymer solutions, the spectra change as a function of both polymer $M_\mathrm{w}$ and concentration. As the polymer concentration increases, the transverse peak shifts to a higher wavelength of 550 nm and the longitudinal peak shifts to a lower wavelength of 675 nm. At the highest concentrations of low $M_\mathrm{w}$ polymer, the longitudinal peak broadens so significantly as to nearly disappear. Furthermore, the relative concentration of polymer needed to induce these changes in optical properties becomes progressively larger for polymers of greater $M_\mathrm{w}$. 

Previous studies have associated similar changes in optical spectra to the aggregation of AuNRs.\cite{Tam2010,Jiang2012,Chen2013,Wang2014} Because the grafted PEO chains have a low $M_\mathrm{W}$, the dissolved chains with much higher $M_\mathrm{W}$ do not significantly penetrate the grafted layer. \cite{Stiakakis2002, Wilk2010, Poling-Skutvik2017} Hence, depletion attractions are expected to arise as the dissolved PEO is excluded from a region near the AuNR surface. The changes to the optical spectra confirm that the depletion attractions indeed induce the aggregation of the nanorods. \EDITS{Resuspension of the aggregated AuNRs in water recovers the original optical absorption spectra (Supporting Information, Fig. S1), indicating that the aggregation is reversible, as expected for entropic depletion interactions.} \EDITS{To support the optical absorption measurements, we obtain quantitative and semi-quantitative information on the structure of the AuNR aggregates using small angle x-ray scattering.} 

\subsection{Structural characterization of AuNR aggregates}

\begin{figure}[ht!]
\includegraphics[width = 2.9 in]{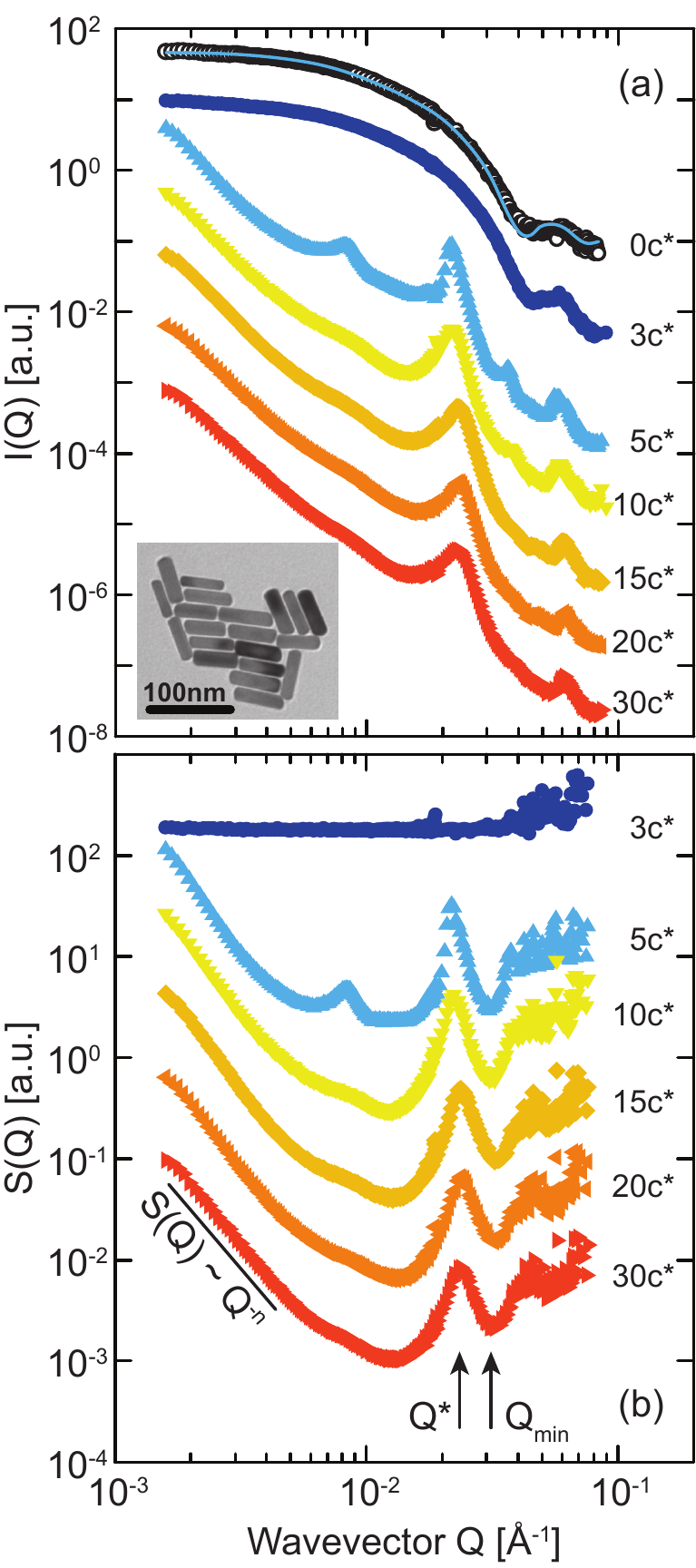}
\caption{\label{fig:IQSQ} (a) SAXS intensity $I(Q)$ as a function of wavevector $Q$ for AuNRs suspended in aqueous solutions of 100 kDa PEO at various concentrations. Solid curve is fit to cylinder form factor (Eq.\ \ref{eqn:cylinder_form_factor}). \textit{Inset:} TEM micrograph of AuNRs \EDITS{grafted with 2 kDa PEO}. (b) Structure factor $S(Q)$ as a function of wavevector $Q$ for the same samples. Solid line indicates the low-$Q$ slope $n$ and arrows indicate wavevectors corresponding to a maximum in $S(Q)$ at $Q^*$ and the first minimum after maximum at $Q_\mathrm{min}$. All data are shifted vertically for clarity. \EDITS{SAXS data for remaining depletants shown in Supporting Information.}}
\end{figure}

Using SAXS, we assess the structural properties of the AuNR aggregates in these polymer solutions, for which the nanoscale polymer structural length scales are comparably sized to the dimensions of the AuNRs. \EDITS{Although the nanorods are geometrically anisotropic, they orient randomly within the suspension to generate azimuthally isotropic scattering patterns.} The \EDITS{azimuthally averaged 1-D} SAXS scattering intensity $I(Q)$ exhibits significant changes with increasing polymer concentration (Fig.\ \ref{fig:IQSQ}). The SAXS intensity for AuNRs dispersed in pure water is well-described by a 1-dimensional cylindrical form factor\cite{Guinier1955} 
\begin{equation}
P(Q) = A\int_0^{\pi/2} f^2(Q,\alpha) \sin(\alpha) d\alpha + I_\mathrm{bkg}
\label{eqn:cylinder_form_factor}
\end{equation}
where
\begin{equation}
f(Q,\alpha) = \frac{\sin (QL\cos(\alpha)/2)}{QL \cos(\alpha)/2} \frac{J_1(Qd_\mathrm{NP}\sin(\alpha)/2)}{Qd_\mathrm{NP}\sin (\alpha)/2},
\end{equation}
$A$ is a prefactor, $\alpha$ is the angle between the cylinder axis and the wavevector $Q$, $J_1$ is the first-order Bessel function of the first kind, and $I_\mathrm{bkg}$ is a constant background scattering. With this model, the AuNR dimensions are $L = 58 \pm 1$ nm and $d_\mathrm{NP} = 19 \pm 1$ nm, in good agreement with the dimensions measured using TEM. At low polymer concentrations (\latin{i.e.} $3c^*$ of 100 kDa PEO), the AuNRs remain well dispersed. At higher concentrations, however, the AuNRs aggregate, leading to  significant scattering intensity \emph{between} AuNRs and the appearance of a structure factor $S(Q)$ (Fig.\ \ref{fig:IQSQ}(b)). \EDITS{Even for aggregated samples, the scattering pattern remains isotropic so that $S(Q) = I(Q)/P(Q)$.} The $Q$-dependence of $S(Q)$ contains information about the structure of the aggregates: the primary peak at $Q^*$ is related to the center-to-center distance $d = 2\pi Q^{*-1}$ between nanorods in an aggregate; the ratio $S(Q^*)/S(Q_\mathrm{min})$ of the structure factor intensity at the primary peak to the first minimum after the peak semi-quantitatively captures the number of nearest neighbors within the aggregate and describes the ordering of the nanoparticles; and the slope $n$ at low-$Q$ captures the mesoscopic fractal dimension of the aggregates.

\begin{figure}[ht!]
\includegraphics[width = 2.9 in]{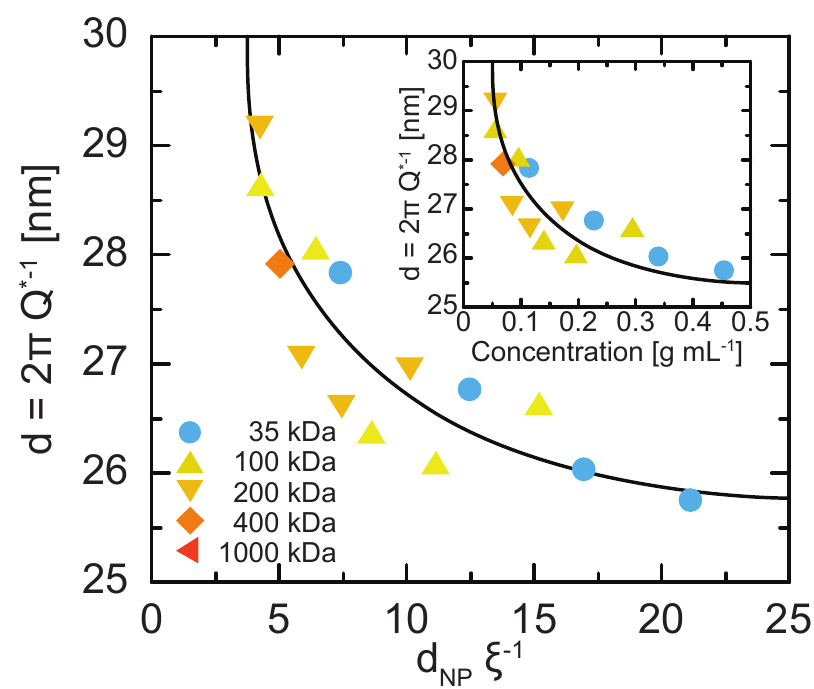}
\caption{\label{fig:dspacing} Interparticle distance $d$ as a function of ratio of particle diameter to correlation length $d_\mathrm{NP}\:\xi^{-1}$ and (\textit{inset}) polymer concentration for various PEO molecular weights. Solid curves are guides to the eye.}
\end{figure}

First, we examine how the center-to-center distance between nanorods changes as a function of depletant $M_\mathrm{w}$ and concentration (Fig.\ \ref{fig:dspacing}). This interparticle distance is a function of the nanorod diameter and the thickness $h$ of the grafted layer according to $d = d_\mathrm{NP} + 2h$. At the lowest concentration of polymer that induces aggregation (e.g.\ 5 $c^*$ of 100 kDa PEO where $d_\mathrm{NP}\: \xi^{-1} \approx 5$), the interparticle distance $d \approx 29$ nm corresponds to cylinders that are packed parallel to each other, as expected from the directionality of the depletion attractions. The resulting grafted thickness at this polymer concentration, $h \approx 5$ nm, is between the size of a free chain in good solvent $R_\mathrm{g} \approx 1.6$ nm and the contour length $l \approx 13.5$ nm for 2 kDa PEO. As expected for dense polymer brushes, the grafted PEO chains are extended beyond their ideal Gaussian configuration and generate a steric repulsion between the AuNRs. For the 5$c^*$ 100 kDa sample, a second peak in $S(Q)$ appears at $Q \approx 0.009$ \AA$^{-1}$, which approximately corresponds to the length of the nanorods and suggests a hierarchical aggregate structure. Because it does not appear for the other solutions, we do not analyze this peak further. As the ratio $d_\mathrm{NP}/\xi$ increases, the interparticle distance monotonically decreases, indicating that the grafted layers are more compressed at higher polymer concentrations. Similar compressions have been observed for polymer-grafted spherical nanoparticles\cite{Poling-Skutvik2017} or star polymers\cite{Stiakakis2002,Wilk2010} dispersed in solutions of linear polymers. Although analytical expressions for the strength of attraction exist in the spherical colloidal limit,\cite{Lekkerkerker2011} no such expressions exist for anisotropic particles in the protein limit. Nevertheless, we expect that the inverse relationship between depletion strength and $\xi$ holds. Thus, we attribute the decrease in interparticle distance with decreasing $\xi$ to an increase in the strength of the depletion attraction. Additionally, the strength of attraction appears to be dependent only on $\xi$ and independent of depletant $M_\mathrm{w}$. 

\begin{figure}[ht!]
\includegraphics[width = 2.9 in]{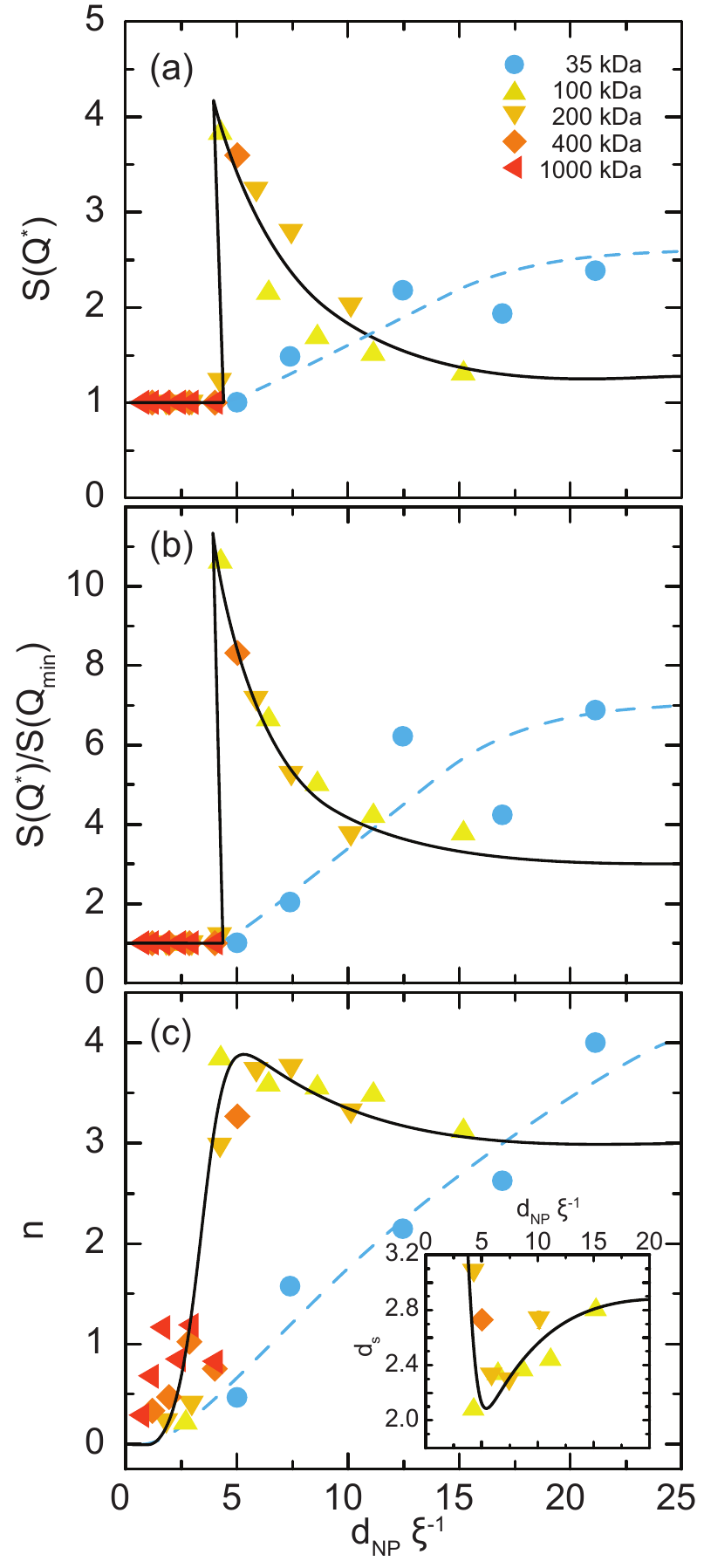}
\caption{\label{fig:SlopeNN} (a) Height of primary peak in $S(Q)$ at $Q = Q^*$, (b) ratio of peak height to first minimum $S(Q^*)/S(Q_\mathrm{min})$, and (c) the low-$Q$ slope $n$ of the SAXS structure factor as a function of the ratio of nanorod diameter to correlation length $d_\mathrm{NP}\: \xi^{-1}$ for various PEO molecular weights. \textit{Inset to (c)}: Surface fractal dimension $d_\mathrm{S} = 6 - n$ as a function of size ratio $d_\mathrm{NP}\: \xi^{-1}$. Dashed blue curves are guides to the eye for the 35 kDa samples. Black curves are guides to the eye for other molecular weights.}
\end{figure}

Although the interparticle distance changes monotonically with $\xi$, the other two structural measures exhibit pronounced non-monotonic trends (Fig.\ \ref{fig:SlopeNN}). Previous reports have shown that the height of the primary peak $S(Q^*)$ increases concomitant with the average number of nearest neighbors.\cite{Ramsay1983,Svensson1980} \EDITS{As the primary peak height $S(Q^*)$ increases, the following minimum $S(Q_\mathrm{min})$ deepens so that the ratio $S(Q^*)/S(Q_\mathrm{min})$ increases \EDITSJCC{concomitant} with \EDITSJCC{the} number of nearest neighbors. Hence,} $S(Q^*)/S(Q_\mathrm{min})$ is a semi-quantitative measure of the number of nearest neighbors, and thus the local number density, within an aggregate.  $S(Q)$ is required to approach 1 as $Q \to \infty$. This high-$Q$ limit is not achieved experimentally, however, so estimating the structure factor peak height introduces an arbitrary vertical scaling factor. Alternatively, the number of nearest neighbors can be quantified by converting $S(Q)$ to the pair distribution function $g(r)$ using an inverse Fourier transform and then integrating over the primary peak.\cite{Salmon1988,Cristiglio2009} Reproducing the primary peak in $g(r)$ from such an inversion, however, requires highly accurate measurements at large $Q$, which are absent for this system. \EDITS{Precisely quantifying the number of nearest neighbors could also be achieved by fitting to an explicit functional form for $S(Q)$, which is common practice for hard sphere suspensions, but such an explicit functional form does not exist for suspensions of anisotropic particles with strong interparticle interactions.} Thus, although $S(Q^*)/S(Q_\mathrm{min})$ is a semi-quantitative measure of the number of nearest neighbors, this metric removes error associated with vertically scaling or inverting $S(Q)$ and hence provides a more accurate understanding of how the local number density varies with $\xi$. 

The height of the primary peak in $S(Q)$ (Fig.\ \ref{fig:SlopeNN}(a)) trends similarly to the ratio $S(Q^*)/S(Q_\mathrm{min})$ (Fig.\ \ref{fig:SlopeNN}(b)) as a function of $d_\mathrm{NP} \: \xi^{-1}$. We focus on the changes to the ratio $S(Q^*)/S(Q_\mathrm{min})$, which exhibits a cleaner collapse due to removal of arbitrary vertical scaling. For $d_\mathrm{NP}\:\xi^{-1} \lesssim 5$, the absence of a local maximum in $S(Q)$ indicates that the AuNRs remain dispersed in solution; thus $S(Q^*)/S(Q_\mathrm{min}) = 1$. At higher polymer concentrations, how $S(Q^*)/S(Q_\mathrm{min})$ varies with $d_\mathrm{NP}\:\xi^{-1}$ depends on the depletant $M_\mathrm{w}$ (Fig.\ \ref{fig:SlopeNN}(b)). For the 35 kDa depletant ($R_\mathrm{g} < d_\mathrm{NP}$), the ratio $S(Q^*)/S(Q_\mathrm{min})$ increases with $d_\mathrm{NP}\:\xi^{-1}$, indicating that the number of nearest neighbors in the aggregate increases with increasing depletion attraction strength. By stark contrast, the ratio $S(Q^*)/S(Q_\mathrm{min})$ for solutions with larger $M_\mathrm{w}$ depletants ($R_\mathrm{g} \gtrsim d_\mathrm{NP}$) discontinuously increases by an order of magnitude when $d_\mathrm{NP}\: \xi^{-1} \approx 5$ and then decreases as $d_\mathrm{NP}\: \xi^{-1}$ is further increased. For these high $M_\mathrm{w}$ depletants, the number of nearest neighbors is maximal at the lowest depletion strength that can still induce aggregation. As the strength of attraction increases, the number of nearest neighbors decreases independently of depletant $M_\mathrm{w}$.

Similar to how the number of nearest neighbors depends on $d_\mathrm{NP}\:\xi^{-1}$, the slope $n$ of $S(Q)$ at low-$Q$ changes monotonically for the 35 kDa depletant and non-monotonically for the higher $M_\mathrm{w}$ depletants (Fig.\ \ref{fig:SlopeNN}(c)). At low concentrations, $n$ is clustered between 0 and 1, indicating that there is little to no structuring on long length scales; the absence of a peak in $S(Q)$ at these concentrations indicates that any structures that are formed do not have controlled interparticle spacing. \EDITS{Due to the differences in polymer-particle interactions when the depletant $R_\mathrm{g}$ is greater than or less than $d_\mathrm{NP}$, the fractal dimensions of the assembled structures vary with depletant $M_\mathrm{W}$.} For the 35 kDa depletant, $n$ increases with increasing  polymer concentration, indicating that the resulting aggregates are increasingly dense mass fractals. For higher $M_\mathrm{w}$ depletants, $n$ attains a maximum when $d_\mathrm{NP}\: \xi^{-1} \approx 5$ and ranges from 3 to 4 for all solutions that induce aggregation\EDITS{, indicating that these depletants induce surface fractal structures.\cite{Schaefer1986,Hurd1987}} Thus, $n$ for the higher $M_\mathrm{w}$ depletants can be related to a surface fractal dimension $d_\mathrm{s} = 6 - n$ (inset to Fig.\ \ref{fig:SlopeNN}(c)), which ranges in value from 2 to 3 corresponding to structures with smooth (\emph{i.e.} dense) and rough interfaces, respectively.\cite{Sinha1989,Radlinski1999}. Based on this change in fractal dimension, the aggregates formed with high $M_\mathrm{w}$ depletants are densest when $d_\mathrm{NP}\: \xi^{-1} \approx 5$ and are increasingly rough as the polymer concentration increases. 

The non-monotonic behavior of the fractal dimension and the number of nearest neighbors suggests that the aggregate structure is the result of two opposing kinetic processes. Aggregation in the colloidal size limit depends on the rate and strength with which two colloids bind upon contact and the rate at which they transport towards each other.\cite{Weitz1985,Lin1989} When the bonds are strong, the rate of diffusive transport limits the growth kinetics of the aggregate, termed diffusion-limited aggregation (DLA).\cite{Witten1981} In the opposite extreme of weak bonds and fast diffusive transport, the rate of the reaction binding the colloids together restricts the aggregate growth, resulting in reaction-limited aggregation (RLA).\cite{Ball1987} RLA aggregates are denser with a larger fractal dimension than those formed by DLA because the colloids within an aggregate are able to rearrange into a more energetically favorable configuration. The strong bonds present in DLA prevent this rearrangement. Although these pictures were originally developed in the colloidal limit with small depletants, we posit that the competition between transport and reaction also shapes depletion-induced aggregates in the protein limit.

\section{Discussion}

The balance between the rate of irreversible binding and the rate of transport depends on the strength of the depletion interactions and the nanoparticle diffusivity. By quantifying the monotonic change in interparticle distance (Fig.\ \ref{fig:dspacing}), we show that the depletion interaction in the ``protein limit" increases in strength as $\xi$ decreases. By contrast, the diffusive transport rate through polymer solutions is expected to decrease along with $\xi$. Whereas colloidal dynamics couple to the bulk viscoelasticity of the surrounding fluid,\cite{Squires2010} nanoparticle dynamics depend sensitively on the nanoparticle size and an effective local viscosity.\cite{Cai2011,Poling-Skutvik2015} When the diameter of a spherical nanoparticle is comparable to the polymer $R_\mathrm{g}$ or $\xi$ in solution, the diffusivity decays as a function of $d_\mathrm{NP}\: \xi^{-1}$, although the functional form may be exponential based on hydrodynamic arguments\cite{Cukier1984,Phillies1985,Cheng2002} or a power law based on scaling arguments.\cite{Cai2011} Nevertheless, it is well established that the diffusive transport of nanoparticles in polymer solutions depends solely on $d_\mathrm{NP}\: \xi^{-1}$. Thus, the two processes controlling aggregation of AuNRs when $d_\mathrm{NP} < R_\mathrm{g}$ are both functions only of $d_\mathrm{NP}\, \xi^{-1}$ but scale in opposite directions: the strength of the depletion attractions increase and the AuNR dynamics decrease with decreasing $\xi$. 

\begin{figure}[ht!]
\includegraphics[width = 3.25 in]{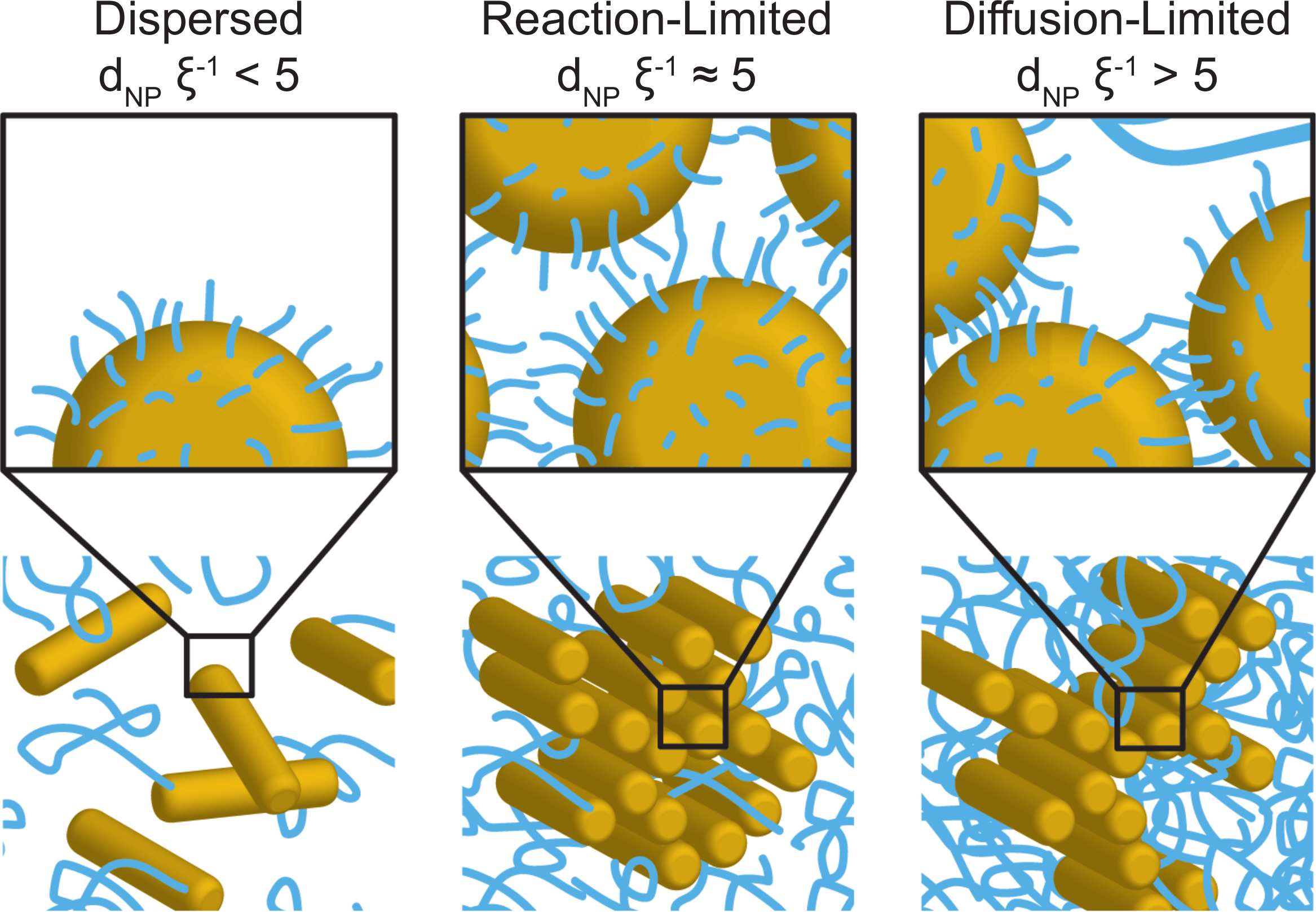}
\caption{\label{fig:schematic} Illustration of the proposed structures formed by the AuNRs in polymer solutions: (\textit{left}) individually dispersed rods at low polymer concentrations, (\textit{middle}) dense RLA aggregates with extended grafted brushes at intermediate concentrations, and (\textit{right}) diffuse DLA aggregates with compressed grafted brushes at high concentrations. }
\end{figure}

To synthesize the changes in various structural properties investigated here, we propose that the aggregation process transitions from RLA to DLA as the depletant concentration is increased. This transition results in the structural assemblies of AuNRs illustrated in Fig.\ \ref{fig:schematic}. At low polymer concentrations ($d_\mathrm{NP}\: \xi^{-1} < 5$), the depletion attractions are too weak to induce aggregation and the AuNRs remain individually dispersed in solution. With the addition of a small amount of polymer so that $d_\mathrm{NP} \,\xi^{-1} \approx 5$, the depletion attractions become strong enough to induce aggregation but the AuNR dynamics are still fast compared to the rate of irreversible binding. Thus, the AuNRs form dense aggregates via RLA with a high number of nearest neighbors and extended grafted layers. At higher concentrations of dissolved polymer ($d_\mathrm{NP}\: \xi^{-1} > 5$), stronger attraction forces and slower dynamics lead to DLA; the resulting aggregates are less dense, each AuNR has fewer nearest neighbors than for aggregates formed at lower concentrations, and the grafted layers are compressed. 

The collapse of the structural properties onto a master curve as a function of $d_\mathrm{NP}\:\xi^{-1}$ in a depletant system is a unique feature of the ``protein limit." In this limit, the bond strength and dynamics are both functions of $\xi$. By contrast, in the traditional colloidal limit the particle dynamics do not depend on $\xi$ but on the bulk viscosity,\cite{Squires2010} which in turn depends on the relative polymer concentration $c/c^*$ (Fig.\ \ref{fig:Rheology}). The collapse of structural properties requires $d_\mathrm{NP}$ to be smaller than the depletant $R_\mathrm{g}$. For the 35 kDa depletant, the nanorod diameter is larger than the depletant $R_\mathrm{g}$ and the resulting aggregate structures are markedly different. The origin of this difference is unclear but may lie in a change in the range of attraction,\cite{Poon1998,Zhang2013} or in different transport properties of the AuNRs when $d_\mathrm{NP} \gtrsim R_\mathrm{g}$.\cite{Cai2011} 

The assembly of nanomaterials into larger scale structures in complex environments is a physically rich problem. By using polymer-induced depletion in the ``protein limit'' and functionalizing the nanoparticle interface with neutrally interacting polymer chains, we isolate the effect of depletant concentration and $M_\mathrm{w}$ on structural properties from complicating factors such as charge stabilization or particle-depletant interactions. While this simplified model system allows us to quantify the structural changes that occur during assembly, other systems may have different characteristics that affect structural properties. For example, in the limit of large $M_\mathrm{w}$ or low grafting density of the grafted polymer, dissolved polymers may penetrate the grafted layer and thus no longer induce depletion attractions.\cite{Stiakakis2002} Additionally, changing the interactions or surface functionalization of the AuNRs will modify the depletion interactions. Attractive interactions between dissolved PEO and the particle surface may greatly reduce the effects of depletion.\cite{Carvalho1993,Zhang2012} For a charged surface, the long-range electrostatic repulsions will shift the aggregation threshold to higher polymer concentrations, potentially rendering the RLA region inaccessible. These types of interactions, along with particle size and shape, may shift the critical ratio of $d_\mathrm{NP} \, \xi^{-1} \approx 5$ at which the depletants induce aggregation. 
\EDITS{For colloids, the strength of depletion attractions scales proportionally with colloid diameter\EDITSJCC{.}\cite{Lekkerkerker2011} \EDITSJCC{We} expect \EDITSJCC{this scaling} to hold for these anisotropic nanorods. \EDITSJCC{Whereas} the structural properties collapse as a function of $d_\mathrm{NP} \, \xi^{-1}$, the role of nanorod length is likely more complicated. In the spherical limit where $L = d_\mathrm{NP}$, the depletion interactions \EDITSJCC{are} isotropic and local anisotropic assembly \EDITSJCC{cannot be achieved}. In the opposite limit of infinitely large rods $L \gg d_\mathrm{NP}$, the rods are kinetically trapped by the surrounding polymer matrix,\cite{Chatterjee2008,Fakhri2010} \EDITSJCC{and thus cannot reorient to align in parallel.} Between these limits, we expect the physics described here to hold, resulting in a transition between RLA and DLA structures. For longer rods, the depletion interactions should be stronger because of an increase in overlap volume with increasing length, thus shifting the critical $d_\mathrm{NP} \, \xi^{-1}$ ratio to lower values.}
Finally, because the aggregation process is kinetically controlled and therefore path dependent, solution preparation and processing will likely affect the nanorod structure. As one example, if instead of dispersing the AuNRs directly into polymer solutions the polymer is slowly added to the AuNR suspension, the DLA structures may not form. Although determining the effects of each of these parameters is beyond the scope of this study, they are relevant to many applications and should be considered when targeting specific structures. Nevertheless, this study describes how to prepare nanoparticle assembles \emph{in situ} and characterize their structural properties via scattering. This approach has broad potential for guiding the design of fractal assemblies of anisotropic nanoparticles and their properties in practically-relevant environmental conditions.

\section{Conclusions}

The assembly of anisotropic gold nanoparticles into fractal structures with controlled properties is essential for applications in drug delivery, theranostics, sensing, catalysis, and electronics. Here, we demonstrate control over the assembly of anisotropic AuNRs using polymer-induced depletion attractions. Nanorods functionalized with short PEO polymer chains are stable in aqueous suspensions and neutrally interact with dissolved PEO polymers. Because the nanorods and polymers are comparably sized, the depletion forces and transport rates are dependent on the length scale ratio $d_\mathrm{NP} \, \xi^{-1}$. The physics controlling the assembly process transitions from reaction-limited at low polymer concentrations (low $d_\mathrm{NP} \, \xi^{-1}$)  to diffusion-limited at high polymer concentrations (high $d_\mathrm{NP} \, \xi^{-1}$), independent of depletant molecular weight. We exploit this change in physics to form structures with controlled interparticle spacings, number density, and fractal dimension. 

The production of bulk materials with controlled nanoscale properties limits the deployment of nanotechnology in many applications. \EDITS{Porous nanomaterials demonstrate unique mechnical properties while remaining lightweight,\cite{Meza2014} but production of these materials often require complex and expensive fabrication techniques.} Using depletion interactions in the ``protein limit'' to assemble anisotropic nanoparticles is a facile route to produce open, porous materials \EDITSJCC{in bulk and} \emph{in situ} with tunable nanoscale properties. The decoupling of structural properties from the polymer $M_\mathrm{w}$ enables the bulk mechanical moduli and nanoscale structures of polymer composites to be independently tuned; varying the molecular weight and grafting density of the grafted polymer grants additional control over the interparticle spacing. Beyond the development of novel materials, the assembly of gold nanoparticles has proven to be beneficial for \EDITS{biosensing\cite{Cao2014}} and targeted drug delivery by reducing the rate of exocytosis.\cite{Nam2009} \EDITS{In addition to generating controlled structures, this study elucidates physics that will \EDITSJCC{improve understanding of the behavior of} polymer-grafted nanomaterials in crowded media.} Biopolymers and proteins present at volume fractions of up to 40\% crowd nanoparticles deployed in biological environments. With \emph{a priori} knowledge of the density or effective correlation length in the cellular cytoplasm or other biological fluids, the nanoparticle dimensions can be varied to generate structures \emph{in vivo} with desired properties.  

\begin{suppinfo}
The supporting information provides details on the \EDITS{aggregate reversibility, correlation lengths as a function of depletant $M_\mathrm{W}$ and concentration, and additional SAXS data.}\end{suppinfo}

\begin{acknowledgement}
We thank Nayoung Park for helpful discussions. This research used resources of the Advanced Photon Source, a U.S. Department of Energy (DOE) Office of Science User Facility operated for the DOE Office of Science by Argonne National Laboratory under Contract No. DE-AC02-06CH11357. JCC acknowledges support from the Welch Foundation (E-1869).

\end{acknowledgement}

\bibliography{biblio}

\providecommand{\latin}[1]{#1}
\makeatletter
\providecommand{\doi}
  {\begingroup\let\do\@makeother\dospecials
  \catcode`\{=1 \catcode`\}=2 \doi@aux}
\providecommand{\doi@aux}[1]{\endgroup\texttt{#1}}
\makeatother
\providecommand*\mcitethebibliography{\thebibliography}
\csname @ifundefined\endcsname{endmcitethebibliography}
  {\let\endmcitethebibliography\endthebibliography}{}
\begin{mcitethebibliography}{82}
\providecommand*\natexlab[1]{#1}
\providecommand*\mciteSetBstSublistMode[1]{}
\providecommand*\mciteSetBstMaxWidthForm[2]{}
\providecommand*\mciteBstWouldAddEndPuncttrue
  {\def\EndOfBibitem{\unskip.}}
\providecommand*\mciteBstWouldAddEndPunctfalse
  {\let\EndOfBibitem\relax}
\providecommand*\mciteSetBstMidEndSepPunct[3]{}
\providecommand*\mciteSetBstSublistLabelBeginEnd[3]{}
\providecommand*\EndOfBibitem{}
\mciteSetBstSublistMode{f}
\mciteSetBstMaxWidthForm{subitem}{(\alph{mcitesubitemcount})}
\mciteSetBstSublistLabelBeginEnd
  {\mcitemaxwidthsubitemform\space}
  {\relax}
  {\relax}

\bibitem[Alkilany and Murphy(2010)Alkilany, and Murphy]{Alkilany2010}
Alkilany,~A.~M.; Murphy,~C.~J. {Toxicity and Cellular Uptake of Gold
  Nanoparticles: What We have Learned so Far?} \emph{J. Nanoparticle Res.}
  \textbf{2010}, \emph{12}, 2313--2333\relax
\mciteBstWouldAddEndPuncttrue
\mciteSetBstMidEndSepPunct{\mcitedefaultmidpunct}
{\mcitedefaultendpunct}{\mcitedefaultseppunct}\relax
\EndOfBibitem
\bibitem[Choi \latin{et~al.}(2011)Choi, Kim, Kang, Byeon, Kim, and
  Tae]{Choi2011}
Choi,~W.~I.; Kim,~J.~Y.; Kang,~C.; Byeon,~C.~C.; Kim,~Y.~H.; Tae,~G. {Tumor
  Regression in vivo by Photothermal Therapy Based on Gold-Nanorod-Loaded,
  Functional Nanocarriers}. \emph{ACS Nano} \textbf{2011}, \emph{5},
  1995--2003\relax
\mciteBstWouldAddEndPuncttrue
\mciteSetBstMidEndSepPunct{\mcitedefaultmidpunct}
{\mcitedefaultendpunct}{\mcitedefaultseppunct}\relax
\EndOfBibitem
\bibitem[Saha \latin{et~al.}(2012)Saha, Agasti, Kim, Li, and Rotello]{Saha2012}
Saha,~K.; Agasti,~S.~S.; Kim,~C.; Li,~X.; Rotello,~V.~M. Gold Nanoparticles in
  Chemical and Biological Sensing. \emph{Chem. Rev.} \textbf{2012}, \emph{112},
  2739--2779\relax
\mciteBstWouldAddEndPuncttrue
\mciteSetBstMidEndSepPunct{\mcitedefaultmidpunct}
{\mcitedefaultendpunct}{\mcitedefaultseppunct}\relax
\EndOfBibitem
\bibitem[Romo-Herrera \latin{et~al.}(2011)Romo-Herrera, Alvarez-Puebla, and
  Liz-Marz{\'{a}}n]{Romo-Herrera2011}
Romo-Herrera,~J.~M.; Alvarez-Puebla,~R.~A.; Liz-Marz{\'{a}}n,~L.~M. {Controlled
  Assembly of Plasmonic Colloidal Nanoparticle Clusters}. \emph{Nanoscale}
  \textbf{2011}, \emph{3}, 1304--1315\relax
\mciteBstWouldAddEndPuncttrue
\mciteSetBstMidEndSepPunct{\mcitedefaultmidpunct}
{\mcitedefaultendpunct}{\mcitedefaultseppunct}\relax
\EndOfBibitem
\bibitem[Thompson(2007)]{Thompson2007}
Thompson,~D.~T. Using Gold Nanoparticles for Catalysis. \emph{Nano Today}
  \textbf{2007}, \emph{2}, 40 -- 43\relax
\mciteBstWouldAddEndPuncttrue
\mciteSetBstMidEndSepPunct{\mcitedefaultmidpunct}
{\mcitedefaultendpunct}{\mcitedefaultseppunct}\relax
\EndOfBibitem
\bibitem[Orendorff \latin{et~al.}(2005)Orendorff, Hankins, and
  Murphy]{Orendorff2005}
Orendorff,~C.~J.; Hankins,~P.~L.; Murphy,~C.~J. {pH-Triggered Assembly of Gold
  Nanorods}. \emph{Langmuir} \textbf{2005}, \emph{21}, 2022--2026\relax
\mciteBstWouldAddEndPuncttrue
\mciteSetBstMidEndSepPunct{\mcitedefaultmidpunct}
{\mcitedefaultendpunct}{\mcitedefaultseppunct}\relax
\EndOfBibitem
\bibitem[Nam \latin{et~al.}(2009)Nam, Won, Jin, Chung, and Kim]{Nam2009}
Nam,~J.; Won,~N.; Jin,~H.; Chung,~H.; Kim,~S. {pH-Induced Aggregation of Gold
  Nanoparticles for Photothermal Cancer Therapy}. \emph{J. Am. Chem. Soc.}
  \textbf{2009}, \emph{131}, 13639--13645\relax
\mciteBstWouldAddEndPuncttrue
\mciteSetBstMidEndSepPunct{\mcitedefaultmidpunct}
{\mcitedefaultendpunct}{\mcitedefaultseppunct}\relax
\EndOfBibitem
\bibitem[Boles \latin{et~al.}(2016)Boles, Engel, and Talapin]{Boles2016}
Boles,~M.~A.; Engel,~M.; Talapin,~D.~V. {Self-Assembly of Colloidal
  Nanocrystals: From Intricate Structures to Functional Materials}. \emph{Chem.
  Rev.} \textbf{2016}, \emph{116}, 11220--11289\relax
\mciteBstWouldAddEndPuncttrue
\mciteSetBstMidEndSepPunct{\mcitedefaultmidpunct}
{\mcitedefaultendpunct}{\mcitedefaultseppunct}\relax
\EndOfBibitem
\bibitem[Tam \latin{et~al.}(2010)Tam, Murthy, Ingram, Nguyen, Sokolov, and
  Johnston]{Tam2010}
Tam,~J.~M.; Murthy,~A.~K.; Ingram,~D.~R.; Nguyen,~R.; Sokolov,~K.~V.;
  Johnston,~K.~P. {Kinetic Assembly of Near-IR-Active Gold Nanoclusters Using
  Weakly Adsorbing Polymers to Control the Size}. \emph{Langmuir}
  \textbf{2010}, \emph{26}, 8988--8999\relax
\mciteBstWouldAddEndPuncttrue
\mciteSetBstMidEndSepPunct{\mcitedefaultmidpunct}
{\mcitedefaultendpunct}{\mcitedefaultseppunct}\relax
\EndOfBibitem
\bibitem[Murthy \latin{et~al.}(2013)Murthy, Stover, Borwankar, Nie,
  Gourisankar, Truskett, Sokolov, and Johnston]{Murthy2013}
Murthy,~A.~K.; Stover,~R.~J.; Borwankar,~A.~U.; Nie,~G.~D.; Gourisankar,~S.;
  Truskett,~T.~M.; Sokolov,~K.~V.; Johnston,~K.~P. {Equilibrium Gold
  Nanoclusters Quenched with Biodegradable Polymers}. \emph{ACS Nano}
  \textbf{2013}, \emph{7}, 239--251\relax
\mciteBstWouldAddEndPuncttrue
\mciteSetBstMidEndSepPunct{\mcitedefaultmidpunct}
{\mcitedefaultendpunct}{\mcitedefaultseppunct}\relax
\EndOfBibitem
\bibitem[Albanese and Chan(2011)Albanese, and Chan]{Albanese2011}
Albanese,~A.; Chan,~W. C.~W. {Effect of Gold Nanoparticle Aggregation on Cell
  Uptake and Toxicity}. \emph{ACS Nano} \textbf{2011}, \emph{5},
  5478--5489\relax
\mciteBstWouldAddEndPuncttrue
\mciteSetBstMidEndSepPunct{\mcitedefaultmidpunct}
{\mcitedefaultendpunct}{\mcitedefaultseppunct}\relax
\EndOfBibitem
\bibitem[Schwartzberg \latin{et~al.}(2004)Schwartzberg, Grant, Wolcott, Talley,
  Huser, Bogomolni, and Zhang]{Schwartzberg2004}
Schwartzberg,~A.~M.; Grant,~C.~D.; Wolcott,~A.; Talley,~C.~E.; Huser,~T.~R.;
  Bogomolni,~R.; Zhang,~J.~Z. {Unique Gold Nanoparticle Aggregates as a Highly
  Active Surface-Enhanced Raman Scattering Substrate}. \emph{J. Phys. Chem. B}
  \textbf{2004}, \emph{108}, 19191--19197\relax
\mciteBstWouldAddEndPuncttrue
\mciteSetBstMidEndSepPunct{\mcitedefaultmidpunct}
{\mcitedefaultendpunct}{\mcitedefaultseppunct}\relax
\EndOfBibitem
\bibitem[Paul and Robeson(2008)Paul, and Robeson]{Paul2008}
Paul,~D.~R.; Robeson,~L.~M. {Polymer Nanotechnology: Nanocomposites}.
  \emph{Polymer} \textbf{2008}, \emph{49}, 3187--3204\relax
\mciteBstWouldAddEndPuncttrue
\mciteSetBstMidEndSepPunct{\mcitedefaultmidpunct}
{\mcitedefaultendpunct}{\mcitedefaultseppunct}\relax
\EndOfBibitem
\bibitem[Kumar and Krishnamoorti(2010)Kumar, and Krishnamoorti]{Kumar2010}
Kumar,~S.~K.; Krishnamoorti,~R. {Nanocomposites: Structure, Phase Behavior, and
  Properties}. \emph{Annu. Rev. Chem. Biomol. Eng.} \textbf{2010}, \emph{1},
  37--58\relax
\mciteBstWouldAddEndPuncttrue
\mciteSetBstMidEndSepPunct{\mcitedefaultmidpunct}
{\mcitedefaultendpunct}{\mcitedefaultseppunct}\relax
\EndOfBibitem
\bibitem[Dimon \latin{et~al.}(1986)Dimon, Sinha, Weitz, Safinya, Smith, Varady,
  and Lindsay]{Dimon1986}
Dimon,~P.; Sinha,~S.~K.; Weitz,~D.~A.; Safinya,~C.~R.; Smith,~G.~S.;
  Varady,~W.~A.; Lindsay,~H.~M. {Structure of Aggregated Gold Colloids}.
  \emph{Phys. Rev. Lett.} \textbf{1986}, \emph{57}, 595--598\relax
\mciteBstWouldAddEndPuncttrue
\mciteSetBstMidEndSepPunct{\mcitedefaultmidpunct}
{\mcitedefaultendpunct}{\mcitedefaultseppunct}\relax
\EndOfBibitem
\bibitem[Lin \latin{et~al.}(1989)Lin, Lindsay, Weitz, Ball, Klein, and
  Meakin]{Lin1989}
Lin,~M.~Y.; Lindsay,~H.~M.; Weitz,~D.~A.; Ball,~R.~C.; Klein,~R.; Meakin,~P.
  {Universality in Colloid Aggregation}. \emph{Nature} \textbf{1989},
  \emph{339}, 360--362\relax
\mciteBstWouldAddEndPuncttrue
\mciteSetBstMidEndSepPunct{\mcitedefaultmidpunct}
{\mcitedefaultendpunct}{\mcitedefaultseppunct}\relax
\EndOfBibitem
\bibitem[Boal \latin{et~al.}(2000)Boal, Ilhan, {DeRouchey}, Thurn-Albrecht,
  Russell, and Rotello]{Boal2000}
Boal,~A.~K.; Ilhan,~F.; {DeRouchey},~J.~E.; Thurn-Albrecht,~T.; Russell,~T.~P.;
  Rotello,~V.~M. Self-Assembly of Nanoparticles into Structured Spherical and
  Network Aggregates. \emph{Nature} \textbf{2000}, \emph{404}, 746--748\relax
\mciteBstWouldAddEndPuncttrue
\mciteSetBstMidEndSepPunct{\mcitedefaultmidpunct}
{\mcitedefaultendpunct}{\mcitedefaultseppunct}\relax
\EndOfBibitem
\bibitem[Akcora \latin{et~al.}(2009)Akcora, Liu, Kumar, Moll, Li, Benicewicz,
  Schadler, Acehan, Panagiotopoulos, Pryamitsyn, Ganesan, Ilavsky,
  Thiyagarajan, Colby, and Douglas]{Akcora2009}
Akcora,~P.; Liu,~H.; Kumar,~S.~K.; Moll,~J.; Li,~Y.; Benicewicz,~B.~C.;
  Schadler,~L.~S.; Acehan,~D.; Panagiotopoulos,~A.~Z.; Pryamitsyn,~V.;
  Ganesan,~V.; Ilavsky,~J.; Thiyagarajan,~P.; Colby,~R.~H.; Douglas,~J.~F.
  {Anisotropic Self-Assembly of Spherical Polymer-Grafted Nanoparticles}.
  \emph{Nat. Mater.} \textbf{2009}, \emph{8}, 354--359\relax
\mciteBstWouldAddEndPuncttrue
\mciteSetBstMidEndSepPunct{\mcitedefaultmidpunct}
{\mcitedefaultendpunct}{\mcitedefaultseppunct}\relax
\EndOfBibitem
\bibitem[Kumar \latin{et~al.}(2013)Kumar, Jouault, Benicewicz, and
  Neely]{Kumar2013}
Kumar,~S.~K.; Jouault,~N.; Benicewicz,~B.; Neely,~T. {Nanocomposites with
  Polymer Grafted Nanoparticles}. \emph{Macromolecules} \textbf{2013},
  \emph{46}, 3199--3214\relax
\mciteBstWouldAddEndPuncttrue
\mciteSetBstMidEndSepPunct{\mcitedefaultmidpunct}
{\mcitedefaultendpunct}{\mcitedefaultseppunct}\relax
\EndOfBibitem
\bibitem[Frank \latin{et~al.}(1998)Frank, Poncharal, Wang, and Heer]{Frank1998}
Frank,~S.; Poncharal,~P.; Wang,~Z.~L.; Heer,~W.~A. Carbon Nanotube Quantum
  Resistors. \emph{Science} \textbf{1998}, \emph{280}, 1744--1746\relax
\mciteBstWouldAddEndPuncttrue
\mciteSetBstMidEndSepPunct{\mcitedefaultmidpunct}
{\mcitedefaultendpunct}{\mcitedefaultseppunct}\relax
\EndOfBibitem
\bibitem[Baranov \latin{et~al.}(2010)Baranov, Fiore, {Van Huis}, Giannini,
  Falqui, Lafont, Zandbergen, Zanella, Cingolani, and Manna]{Baranov2010}
Baranov,~D.; Fiore,~A.; {Van Huis},~M.; Giannini,~C.; Falqui,~A.; Lafont,~U.;
  Zandbergen,~H.; Zanella,~M.; Cingolani,~R.; Manna,~L. {Assembly of Colloidal
  Semiconductor Nanorods in Solution by Depletion Attraction}. \emph{Nano
  Lett.} \textbf{2010}, \emph{10}, 743--749\relax
\mciteBstWouldAddEndPuncttrue
\mciteSetBstMidEndSepPunct{\mcitedefaultmidpunct}
{\mcitedefaultendpunct}{\mcitedefaultseppunct}\relax
\EndOfBibitem
\bibitem[Park \latin{et~al.}(2010)Park, Koerner, and Vaia]{Park2010}
Park,~K.; Koerner,~H.; Vaia,~R.~A. {Depletion-Induced Shape and Size Selection
  of Gold Nanoparticles}. \emph{Nano Lett.} \textbf{2010}, \emph{10},
  1433--1439\relax
\mciteBstWouldAddEndPuncttrue
\mciteSetBstMidEndSepPunct{\mcitedefaultmidpunct}
{\mcitedefaultendpunct}{\mcitedefaultseppunct}\relax
\EndOfBibitem
\bibitem[Song \latin{et~al.}(2003)Song, Kinloch, and Windle]{Song2003}
Song,~W.; Kinloch,~I.~A.; Windle,~A.~H. Nematic Liquid Crystallinity of
  Multiwall Carbon Nanotubes. \emph{Science} \textbf{2003}, \emph{302},
  1363\relax
\mciteBstWouldAddEndPuncttrue
\mciteSetBstMidEndSepPunct{\mcitedefaultmidpunct}
{\mcitedefaultendpunct}{\mcitedefaultseppunct}\relax
\EndOfBibitem
\bibitem[Rai \latin{et~al.}(2006)Rai, Pinnick, Parra-Vasquez, Davis, Schmidt,
  Hauge, Smalley, and Pasquali]{Rai2006}
Rai,~P.~K.; Pinnick,~R.~A.; Parra-Vasquez,~A. N.~G.; Davis,~V.~A.;
  Schmidt,~H.~K.; Hauge,~R.~H.; Smalley,~R.~E.; Pasquali,~M. {Isotropic-Nematic
  Phase Transition of Single-Walled Carbon Nanotubes in Strong Acids}. \emph{J.
  Am. Chem. Soc.} \textbf{2006}, \emph{128}, 591--595\relax
\mciteBstWouldAddEndPuncttrue
\mciteSetBstMidEndSepPunct{\mcitedefaultmidpunct}
{\mcitedefaultendpunct}{\mcitedefaultseppunct}\relax
\EndOfBibitem
\bibitem[Zanella \latin{et~al.}(2011)Zanella, Bertoni, Franchini, Brescia,
  Baranov, and Manna]{Zanella2011}
Zanella,~M.; Bertoni,~G.; Franchini,~I.~R.; Brescia,~R.; Baranov,~D.; Manna,~L.
  {Assembly of Shape-Controlled Nanocrystals by Depletion Attraction}.
  \emph{Chem. Commun.} \textbf{2011}, \emph{47}, 203--205\relax
\mciteBstWouldAddEndPuncttrue
\mciteSetBstMidEndSepPunct{\mcitedefaultmidpunct}
{\mcitedefaultendpunct}{\mcitedefaultseppunct}\relax
\EndOfBibitem
\bibitem[Young \latin{et~al.}(2013)Young, Personick, Engel, Damasceno, Barnaby,
  Bleher, Li, Glotzer, Lee, and Mirkin]{Young2013}
Young,~K.~L.; Personick,~M.~L.; Engel,~M.; Damasceno,~P.~F.; Barnaby,~S.~N.;
  Bleher,~R.; Li,~T.; Glotzer,~S.~C.; Lee,~B.; Mirkin,~C.~A. {A Directional
  Entropic Force Approach to Assemble Anisotropic Nanoparticles into
  Superlattices}. \emph{Angew. Chem. Int. Ed.} \textbf{2013}, \emph{52},
  13980--13984\relax
\mciteBstWouldAddEndPuncttrue
\mciteSetBstMidEndSepPunct{\mcitedefaultmidpunct}
{\mcitedefaultendpunct}{\mcitedefaultseppunct}\relax
\EndOfBibitem
\bibitem[Smay \latin{et~al.}(2002)Smay, Cesarano, and Lewis]{Smay2002}
Smay,~J.~E.; Cesarano,~J.; Lewis,~J.~A. {Directed Colloidal Assembly of 3D
  Periodic Structures}. \emph{Langmuir} \textbf{2002}, \emph{18},
  5429--5437\relax
\mciteBstWouldAddEndPuncttrue
\mciteSetBstMidEndSepPunct{\mcitedefaultmidpunct}
{\mcitedefaultendpunct}{\mcitedefaultseppunct}\relax
\EndOfBibitem
\bibitem[Chatterjee \latin{et~al.}(2008)Chatterjee, Jackson, and
  Krishnamoorti]{Chatterjee2008}
Chatterjee,~T.; Jackson,~A.; Krishnamoorti,~R. {Hierarchical Structure of
  Carbon Nanotube Networks}. \emph{J. Am. Chem. Soc.} \textbf{2008},
  \emph{130}, 6934--6935\relax
\mciteBstWouldAddEndPuncttrue
\mciteSetBstMidEndSepPunct{\mcitedefaultmidpunct}
{\mcitedefaultendpunct}{\mcitedefaultseppunct}\relax
\EndOfBibitem
\bibitem[Lekkerkerker and Tuinier(2011)Lekkerkerker, and
  Tuinier]{Lekkerkerker2011}
Lekkerkerker,~H.~N.; Tuinier,~R. \emph{{Colloids and the Depletion
  Interaction}}; Springer Netherlands: Dordrecht, 2011; Vol. 833\relax
\mciteBstWouldAddEndPuncttrue
\mciteSetBstMidEndSepPunct{\mcitedefaultmidpunct}
{\mcitedefaultendpunct}{\mcitedefaultseppunct}\relax
\EndOfBibitem
\bibitem[Asakura and Oosawa(1958)Asakura, and Oosawa]{Asakura1958}
Asakura,~S.; Oosawa,~F. {Interaction between Particles Suspended in Solutions
  of Macromolecules}. \emph{J. Polym. Sci.} \textbf{1958}, \emph{33},
  183--192\relax
\mciteBstWouldAddEndPuncttrue
\mciteSetBstMidEndSepPunct{\mcitedefaultmidpunct}
{\mcitedefaultendpunct}{\mcitedefaultseppunct}\relax
\EndOfBibitem
\bibitem[Kleshchanok \latin{et~al.}(2008)Kleshchanok, Tuinier, and
  Lang]{Kleshchanok2008}
Kleshchanok,~D.; Tuinier,~R.; Lang,~P.~R. {Direct Measurements of
  Polymer-Induced Forces}. \emph{J. Phys. Condens. Matter} \textbf{2008},
  \emph{20}, 073101\relax
\mciteBstWouldAddEndPuncttrue
\mciteSetBstMidEndSepPunct{\mcitedefaultmidpunct}
{\mcitedefaultendpunct}{\mcitedefaultseppunct}\relax
\EndOfBibitem
\bibitem[de~Gennes(1979)]{DeGennes1979}
de~Gennes,~P. {Colloid Suspensions in a Polymer Solution}. \emph{C. R. Seances
  Acad. Sci., Ser. B} \textbf{1979}, \emph{288}, 359--361\relax
\mciteBstWouldAddEndPuncttrue
\mciteSetBstMidEndSepPunct{\mcitedefaultmidpunct}
{\mcitedefaultendpunct}{\mcitedefaultseppunct}\relax
\EndOfBibitem
\bibitem[Semenov and Shvets(2015)Semenov, and Shvets]{Semenov2015}
Semenov,~A.~N.; Shvets,~A.~A. {Theory of Colloid Depletion Stabilization by
  Unattached and Adsorbed Polymers}. \emph{Soft Matter} \textbf{2015},
  \emph{11}, 8863--8878\relax
\mciteBstWouldAddEndPuncttrue
\mciteSetBstMidEndSepPunct{\mcitedefaultmidpunct}
{\mcitedefaultendpunct}{\mcitedefaultseppunct}\relax
\EndOfBibitem
\bibitem[Mutch \latin{et~al.}(2007)Mutch, van Duijneveldt, and
  Eastoe]{Mutch2007}
Mutch,~K.~J.; van Duijneveldt,~J.~S.; Eastoe,~J. {Colloid -- Polymer Mixtures
  in the Protein Limit}. \emph{Soft Matter} \textbf{2007}, \emph{3},
  155--167\relax
\mciteBstWouldAddEndPuncttrue
\mciteSetBstMidEndSepPunct{\mcitedefaultmidpunct}
{\mcitedefaultendpunct}{\mcitedefaultseppunct}\relax
\EndOfBibitem
\bibitem[Kulkarni \latin{et~al.}(1999)Kulkarni, Chatterjee, Schweizer, and
  Zukoski]{Kulkarni1999}
Kulkarni,~A.; Chatterjee,~A.; Schweizer,~K.; Zukoski,~C. {Depletion
  Interactions in the Protein Limit: Effects of Polymer Density Fluctuations}.
  \emph{Phys. Rev. Lett.} \textbf{1999}, \emph{83}, 4554--4557\relax
\mciteBstWouldAddEndPuncttrue
\mciteSetBstMidEndSepPunct{\mcitedefaultmidpunct}
{\mcitedefaultendpunct}{\mcitedefaultseppunct}\relax
\EndOfBibitem
\bibitem[Vivar{\`{e}}s \latin{et~al.}(2002)Vivar{\`{e}}s, Belloni, Tardieu, and
  Bonnet{\'{e}}]{Vivares2002}
Vivar{\`{e}}s,~D.; Belloni,~L.; Tardieu,~A.; Bonnet{\'{e}},~F. {Catching the
  PEG-Induced Attractive Interaction between Proteins.} \emph{Eur. Phys. J. E.
  Soft Matter} \textbf{2002}, \emph{9}, 15--25\relax
\mciteBstWouldAddEndPuncttrue
\mciteSetBstMidEndSepPunct{\mcitedefaultmidpunct}
{\mcitedefaultendpunct}{\mcitedefaultseppunct}\relax
\EndOfBibitem
\bibitem[Chatterjee and Schweizer(1998)Chatterjee, and
  Schweizer]{Chatterjee1998}
Chatterjee,~A.~P.; Schweizer,~K.~S. Microscopic Theory of Polymer-Mediated
  Interactions between Spherical Particles. \emph{J. Chem. Phys.}
  \textbf{1998}, \emph{109}, 10464--10476\relax
\mciteBstWouldAddEndPuncttrue
\mciteSetBstMidEndSepPunct{\mcitedefaultmidpunct}
{\mcitedefaultendpunct}{\mcitedefaultseppunct}\relax
\EndOfBibitem
\bibitem[Wormutht(1991)]{Wormutht1991}
Wormutht,~K.~R. {Patterns of Phase Behavior in Polymer and Amphiphile
  Mixtures}. \emph{Langmuir} \textbf{1991}, \emph{7}, 1622--1626\relax
\mciteBstWouldAddEndPuncttrue
\mciteSetBstMidEndSepPunct{\mcitedefaultmidpunct}
{\mcitedefaultendpunct}{\mcitedefaultseppunct}\relax
\EndOfBibitem
\bibitem[Clegg \latin{et~al.}(1994)Clegg, Williams, Warren, and
  Robb]{Clegg1994}
Clegg,~S.~M.; Williams,~P.~A.; Warren,~P.; Robb,~I.~D. {Phase Behavior of
  Polymers with Concentrated Dispersions of Surfactants}. \emph{Langmuir}
  \textbf{1994}, \emph{10}, 3390--3394\relax
\mciteBstWouldAddEndPuncttrue
\mciteSetBstMidEndSepPunct{\mcitedefaultmidpunct}
{\mcitedefaultendpunct}{\mcitedefaultseppunct}\relax
\EndOfBibitem
\bibitem[Piculell \latin{et~al.}(1996)Piculell, Bergfeldt, and
  Gerdes]{Piculell1996}
Piculell,~L.; Bergfeldt,~K.; Gerdes,~S. {Segregation in Aqueous Mixtures of
  Nonionic Polymers and Surfactant Micelles. Effects of Micelle Size and
  Surfactant Headgroup/Polymer Interactions}. \emph{J. Phys. Chem.}
  \textbf{1996}, \emph{100}, 3675--3679\relax
\mciteBstWouldAddEndPuncttrue
\mciteSetBstMidEndSepPunct{\mcitedefaultmidpunct}
{\mcitedefaultendpunct}{\mcitedefaultseppunct}\relax
\EndOfBibitem
\bibitem[Tuinier \latin{et~al.}(2000)Tuinier, Dhont, and {De
  Kruif}]{Tuinier2000}
Tuinier,~R.; Dhont,~J.~K.; {De Kruif},~C.~G. {Depletion-Induced Phase
  Separation of Aggregated Whey Protein Colloids by an Exocellular
  Polysaccharide}. \emph{Langmuir} \textbf{2000}, \emph{16}, 1497--1507\relax
\mciteBstWouldAddEndPuncttrue
\mciteSetBstMidEndSepPunct{\mcitedefaultmidpunct}
{\mcitedefaultendpunct}{\mcitedefaultseppunct}\relax
\EndOfBibitem
\bibitem[Ramakrishnan \latin{et~al.}(2002)Ramakrishnan, Fuchs, Schweizer, and
  Zukoski]{Ramakrishnan2002a}
Ramakrishnan,~S.; Fuchs,~M.; Schweizer,~K.~S.; Zukoski,~C.~F. {Concentration
  Fluctuations in a Model Colloid-Polymer Suspension: Experimental Tests of
  Depletion Theories}. \emph{Langmuir} \textbf{2002}, \emph{18},
  1082--1090\relax
\mciteBstWouldAddEndPuncttrue
\mciteSetBstMidEndSepPunct{\mcitedefaultmidpunct}
{\mcitedefaultendpunct}{\mcitedefaultseppunct}\relax
\EndOfBibitem
\bibitem[Ramakrishnan \latin{et~al.}(2002)Ramakrishnan, Fuchs, Schweizer, and
  Zukoski]{Ramakrishnan2002b}
Ramakrishnan,~S.; Fuchs,~M.; Schweizer,~K.~S.; Zukoski,~C.~F. {Entropy Driven
  Phase Transitions in Colloid-Polymer Suspensions: Tests of Depletion
  Theories}. \emph{J. Chem. Phys.} \textbf{2002}, \emph{116}, 2201--2212\relax
\mciteBstWouldAddEndPuncttrue
\mciteSetBstMidEndSepPunct{\mcitedefaultmidpunct}
{\mcitedefaultendpunct}{\mcitedefaultseppunct}\relax
\EndOfBibitem
\bibitem[Vliegenthart \latin{et~al.}(2003)Vliegenthart, van Duijneveldt, and
  Vincent]{Vliegenthart2003}
Vliegenthart,~G.~A.; van Duijneveldt,~J.~S.; Vincent,~B. {Phase Transitions and
  Gelation of Silica-Polystyrene Mixtures in Benzene}. \emph{Faraday Discuss.}
  \textbf{2003}, \emph{123}, 65--74\relax
\mciteBstWouldAddEndPuncttrue
\mciteSetBstMidEndSepPunct{\mcitedefaultmidpunct}
{\mcitedefaultendpunct}{\mcitedefaultseppunct}\relax
\EndOfBibitem
\bibitem[Hennequin \latin{et~al.}(2005)Hennequin, Evens, {Quezada Angulo}, and
  {van Duijneveldt}]{Hennequin2005}
Hennequin,~Y.; Evens,~M.; {Quezada Angulo},~C.~M.; {van Duijneveldt},~J.~S.
  {Miscibility of Small Colloidal Spheres with Large Polymers in Good Solvent}.
  \emph{J. Chem. Phys.} \textbf{2005}, \emph{123}, 054906\relax
\mciteBstWouldAddEndPuncttrue
\mciteSetBstMidEndSepPunct{\mcitedefaultmidpunct}
{\mcitedefaultendpunct}{\mcitedefaultseppunct}\relax
\EndOfBibitem
\bibitem[Cai \latin{et~al.}(2011)Cai, Panyukov, and Rubinstein]{Cai2011}
Cai,~L.-H.; Panyukov,~S.; Rubinstein,~M. {Mobility of Nonsticky Nanoparticles
  in Polymer Liquids}. \emph{Macromolecules} \textbf{2011}, \emph{44},
  7853--7863\relax
\mciteBstWouldAddEndPuncttrue
\mciteSetBstMidEndSepPunct{\mcitedefaultmidpunct}
{\mcitedefaultendpunct}{\mcitedefaultseppunct}\relax
\EndOfBibitem
\bibitem[Poling-Skutvik \latin{et~al.}(2015)Poling-Skutvik, Krishnamoorti, and
  Conrad]{Poling-Skutvik2015}
Poling-Skutvik,~R.; Krishnamoorti,~R.; Conrad,~J.~C. {Size-Dependent Dynamics
  of Nanoparticles in Unentangled Polyelectrolyte Solutions}. \emph{ACS Macro
  Lett.} \textbf{2015}, \emph{4}, 1169--1173\relax
\mciteBstWouldAddEndPuncttrue
\mciteSetBstMidEndSepPunct{\mcitedefaultmidpunct}
{\mcitedefaultendpunct}{\mcitedefaultseppunct}\relax
\EndOfBibitem
\bibitem[Ye2(2012)]{Ye2012}
{Improved Size-Tunable Synthesis of Monodisperse Gold Nanorods Through the Use
  of Aromatic Additives}. \emph{ACS Nano} \textbf{2012}, \emph{6},
  2804--2817\relax
\mciteBstWouldAddEndPuncttrue
\mciteSetBstMidEndSepPunct{\mcitedefaultmidpunct}
{\mcitedefaultendpunct}{\mcitedefaultseppunct}\relax
\EndOfBibitem
\bibitem[Harder \latin{et~al.}(1998)Harder, Grunze, Dahint, Whitesides, and
  Laibinis]{Harder1998}
Harder,~P.; Grunze,~M.; Dahint,~R.; Whitesides,~G.~M.; Laibinis,~P.~E.
  {Molecular Conformation in Oligo(ethylene glycol)-Terminated Self-Assembled
  Monolayers on Gold and Silver Surfaces Determines Their Ability To Resist
  Protein Adsorption}. \emph{J. Phys. Chem. B} \textbf{1998}, \emph{102},
  426--436\relax
\mciteBstWouldAddEndPuncttrue
\mciteSetBstMidEndSepPunct{\mcitedefaultmidpunct}
{\mcitedefaultendpunct}{\mcitedefaultseppunct}\relax
\EndOfBibitem
\bibitem[Rahme \latin{et~al.}(2013)Rahme, Chen, Hobbs, Morris, O'Driscoll, and
  Holmes]{Rahme2013}
Rahme,~K.; Chen,~L.; Hobbs,~R.~G.; Morris,~M.~A.; O'Driscoll,~C.; Holmes,~J.~D.
  {PEGylated Gold Nanoparticles: Polymer Quantification as a Function of PEG
  Lengths and Nanoparticle Dimensions}. \emph{RSC Adv.} \textbf{2013},
  \emph{3}, 6085--6094\relax
\mciteBstWouldAddEndPuncttrue
\mciteSetBstMidEndSepPunct{\mcitedefaultmidpunct}
{\mcitedefaultendpunct}{\mcitedefaultseppunct}\relax
\EndOfBibitem
\bibitem[Rubinstein and Colby(2003)Rubinstein, and Colby]{Rubinstein2003}
Rubinstein,~M.; Colby,~R.~H. \emph{{Polymer Physics}}; Oxford University Press:
  New York, 2003\relax
\mciteBstWouldAddEndPuncttrue
\mciteSetBstMidEndSepPunct{\mcitedefaultmidpunct}
{\mcitedefaultendpunct}{\mcitedefaultseppunct}\relax
\EndOfBibitem
\bibitem[Park \latin{et~al.}(2010)Park, Sinha, and Hamad-Schifferli]{Park2010b}
Park,~S.; Sinha,~N.; Hamad-Schifferli,~K. {Effective Size and Zeta Potential of
  Nanorods by {Ferguson} Analysis}. \emph{Langmuir} \textbf{2010}, \emph{26},
  13071--13075\relax
\mciteBstWouldAddEndPuncttrue
\mciteSetBstMidEndSepPunct{\mcitedefaultmidpunct}
{\mcitedefaultendpunct}{\mcitedefaultseppunct}\relax
\EndOfBibitem
\bibitem[Jiang \latin{et~al.}(2012)Jiang, Hore, Gam, and Composto]{Jiang2012}
Jiang,~G.; Hore,~M. J.~A.; Gam,~S.; Composto,~R.~J. {Gold Nanorods Dispersed in
  Homopolymer Films: Optical Properties Controlled by Self-Assembly and
  Percolation of Nanorods}. \emph{ACS Nano} \textbf{2012}, \emph{6},
  1578--1588\relax
\mciteBstWouldAddEndPuncttrue
\mciteSetBstMidEndSepPunct{\mcitedefaultmidpunct}
{\mcitedefaultendpunct}{\mcitedefaultseppunct}\relax
\EndOfBibitem
\bibitem[Chen \latin{et~al.}(2013)Chen, Shao, Li, and Wang]{Chen2013}
Chen,~H.; Shao,~L.; Li,~Q.; Wang,~J. {Gold Nanorods and their Plasmonic
  Properties}. \emph{Chem. Soc. Rev.} \textbf{2013}, \emph{42},
  2679--2724\relax
\mciteBstWouldAddEndPuncttrue
\mciteSetBstMidEndSepPunct{\mcitedefaultmidpunct}
{\mcitedefaultendpunct}{\mcitedefaultseppunct}\relax
\EndOfBibitem
\bibitem[Wang \latin{et~al.}(2014)Wang, Hore, Ye, Zheng, Murray, and
  Composto]{Wang2014}
Wang,~D.; Hore,~M. J.~A.; Ye,~X.; Zheng,~C.; Murray,~C.~B.; Composto,~R.~J.
  {Gold Nanorod Length Controls Dispersion, Local Ordering, and Optical
  Absorption in Polymer Nanocomposite Films}. \emph{Soft Matter} \textbf{2014},
  \emph{10}, 3404--3413\relax
\mciteBstWouldAddEndPuncttrue
\mciteSetBstMidEndSepPunct{\mcitedefaultmidpunct}
{\mcitedefaultendpunct}{\mcitedefaultseppunct}\relax
\EndOfBibitem
\bibitem[Stiakakis \latin{et~al.}(2002)Stiakakis, Vlassopoulos, Likos, Roovers,
  and Meier]{Stiakakis2002}
Stiakakis,~E.; Vlassopoulos,~D.; Likos,~C.~N.; Roovers,~J.; Meier,~G.
  {Polymer-Mediated Melting in Ultrasoft Colloidal Gels}. \emph{Phys. Rev.
  Lett.} \textbf{2002}, \emph{89}, 208302\relax
\mciteBstWouldAddEndPuncttrue
\mciteSetBstMidEndSepPunct{\mcitedefaultmidpunct}
{\mcitedefaultendpunct}{\mcitedefaultseppunct}\relax
\EndOfBibitem
\bibitem[Wilk \latin{et~al.}(2010)Wilk, Hui{\ss}mann, Stiakakis, Kohlbrecher,
  Vlassopoulos, Likos, Meier, Dhont, Petekidis, and Vavrin]{Wilk2010}
Wilk,~A.; Hui{\ss}mann,~S.; Stiakakis,~E.; Kohlbrecher,~J.; Vlassopoulos,~D.;
  Likos,~C.~N.; Meier,~G.; Dhont,~J. K.~G.; Petekidis,~G.; Vavrin,~R. {Osmotic
  Shrinkage in Star/Linear Polymer Mixtures}. \emph{Eur. Phys. J. E}
  \textbf{2010}, \emph{32}, 127--134\relax
\mciteBstWouldAddEndPuncttrue
\mciteSetBstMidEndSepPunct{\mcitedefaultmidpunct}
{\mcitedefaultendpunct}{\mcitedefaultseppunct}\relax
\EndOfBibitem
\bibitem[Poling-Skutvik \latin{et~al.}(2017)Poling-Skutvik, Olafson, Narayanan,
  Stingaciu, Faraone, Conrad, and Krishnamoorti]{Poling-Skutvik2017}
Poling-Skutvik,~R.; Olafson,~K.~N.; Narayanan,~S.; Stingaciu,~L.; Faraone,~A.;
  Conrad,~J.~C.; Krishnamoorti,~R. {Confined Dynamics of Grafted Polymer Chains
  in Solutions of Linear Polymer}. \emph{Macromolecules} \textbf{2017},
  \emph{50}, 7372--7379\relax
\mciteBstWouldAddEndPuncttrue
\mciteSetBstMidEndSepPunct{\mcitedefaultmidpunct}
{\mcitedefaultendpunct}{\mcitedefaultseppunct}\relax
\EndOfBibitem
\bibitem[Guinier and Fournet(1955)Guinier, and Fournet]{Guinier1955}
Guinier,~A.; Fournet,~G. \emph{{Small-Angle Scattering of X-Rays}}; John Wiley
  and Sons: New York, 1955\relax
\mciteBstWouldAddEndPuncttrue
\mciteSetBstMidEndSepPunct{\mcitedefaultmidpunct}
{\mcitedefaultendpunct}{\mcitedefaultseppunct}\relax
\EndOfBibitem
\bibitem[Ramsay and Booth(1983)Ramsay, and Booth]{Ramsay1983}
Ramsay,~J.~D.; Booth,~B.~O. {Determination of Structure in Oxide Sols and Gels
  From Neutron Scattering and Nitrogen Adsorption Measurements}. \emph{J. Chem.
  Soc. Faraday Trans. 1} \textbf{1983}, \emph{79}, 173--184\relax
\mciteBstWouldAddEndPuncttrue
\mciteSetBstMidEndSepPunct{\mcitedefaultmidpunct}
{\mcitedefaultendpunct}{\mcitedefaultseppunct}\relax
\EndOfBibitem
\bibitem[Svensson \latin{et~al.}(1980)Svensson, Sears, Woods, and
  Martel]{Svensson1980}
Svensson,~E.~C.; Sears,~V.~F.; Woods,~A. D.~B.; Martel,~P. {Neutron-Diffraction
  Study of the Static Structure Factor and Pair Correlations in Liquid $^4$He}.
  \emph{Phys. Rev. B} \textbf{1980}, \emph{21}, 3638--3651\relax
\mciteBstWouldAddEndPuncttrue
\mciteSetBstMidEndSepPunct{\mcitedefaultmidpunct}
{\mcitedefaultendpunct}{\mcitedefaultseppunct}\relax
\EndOfBibitem
\bibitem[Salmon(1988)]{Salmon1988}
Salmon,~P. {A Neutron Diffraction Study on the Structure of Liquid Germanium}.
  \emph{J. Phys. F Met. Phys.} \textbf{1988}, \emph{18}, 2345--2352\relax
\mciteBstWouldAddEndPuncttrue
\mciteSetBstMidEndSepPunct{\mcitedefaultmidpunct}
{\mcitedefaultendpunct}{\mcitedefaultseppunct}\relax
\EndOfBibitem
\bibitem[Cristiglio \latin{et~al.}(2009)Cristiglio, Cuello, Piarristeguy, and
  Pradel]{Cristiglio2009}
Cristiglio,~V.; Cuello,~G.~J.; Piarristeguy,~A.~A.; Pradel,~A. {The
  Coordination Number Calculation from Total Structure Factor Measurements}.
  \textbf{2009}, \emph{355}, 1811--1814\relax
\mciteBstWouldAddEndPuncttrue
\mciteSetBstMidEndSepPunct{\mcitedefaultmidpunct}
{\mcitedefaultendpunct}{\mcitedefaultseppunct}\relax
\EndOfBibitem
\bibitem[Schaefer and Keefer(1986)Schaefer, and Keefer]{Schaefer1986}
Schaefer,~D.~W.; Keefer,~K.~D. {Structure of Random Porous Materials: Silica
  Aerogel}. \emph{Phys. Rev. Lett.} \textbf{1986}, \emph{56}, 2199--2202\relax
\mciteBstWouldAddEndPuncttrue
\mciteSetBstMidEndSepPunct{\mcitedefaultmidpunct}
{\mcitedefaultendpunct}{\mcitedefaultseppunct}\relax
\EndOfBibitem
\bibitem[Hurd \latin{et~al.}(1987)Hurd, Schaefer, and Martin]{Hurd1987}
Hurd,~A.~J.; Schaefer,~D.~W.; Martin,~J.~E. {Surface and Mass Fractals in
  Vapor-Phase Aggregates}. \emph{Phys. Rev. A} \textbf{1987}, \emph{35},
  2361--2364\relax
\mciteBstWouldAddEndPuncttrue
\mciteSetBstMidEndSepPunct{\mcitedefaultmidpunct}
{\mcitedefaultendpunct}{\mcitedefaultseppunct}\relax
\EndOfBibitem
\bibitem[Sinha(1989)]{Sinha1989}
Sinha,~S.~K. {Scattering from Fractal Structures}. \emph{Phys. D Nonlinear
  Phenom.} \textbf{1989}, \emph{38}, 310--314\relax
\mciteBstWouldAddEndPuncttrue
\mciteSetBstMidEndSepPunct{\mcitedefaultmidpunct}
{\mcitedefaultendpunct}{\mcitedefaultseppunct}\relax
\EndOfBibitem
\bibitem[Radli{\'{n}}ski \latin{et~al.}(1999)Radli{\'{n}}ski, Radli{\'{n}}ska,
  Agamalian, Wignall, Lindner, and Randl]{Radlinski1999}
Radli{\'{n}}ski,~A.~P.; Radli{\'{n}}ska,~E.~Z.; Agamalian,~M.; Wignall,~G.~D.;
  Lindner,~P.; Randl,~O.~G. {Fractal Geometry of Rocks}. \emph{Phys. Rev.
  Lett.} \textbf{1999}, \emph{82}, 3078--3081\relax
\mciteBstWouldAddEndPuncttrue
\mciteSetBstMidEndSepPunct{\mcitedefaultmidpunct}
{\mcitedefaultendpunct}{\mcitedefaultseppunct}\relax
\EndOfBibitem
\bibitem[Weitz \latin{et~al.}(1985)Weitz, Huang, Lin, and Sung]{Weitz1985}
Weitz,~D.~A.; Huang,~J.~S.; Lin,~M.~Y.; Sung,~J. {Limits of the Fractal
  Dimension for Irreversible Kinetic Aggregation of Gold Colloids}. \emph{Phys.
  Rev. Lett.} \textbf{1985}, \emph{54}, 1416--1419\relax
\mciteBstWouldAddEndPuncttrue
\mciteSetBstMidEndSepPunct{\mcitedefaultmidpunct}
{\mcitedefaultendpunct}{\mcitedefaultseppunct}\relax
\EndOfBibitem
\bibitem[Witten and Sander(1981)Witten, and Sander]{Witten1981}
Witten,~T.~A.; Sander,~L.~M. Diffusion-Limited Aggregation, a Kinetic Critical
  Phenomenon. \emph{Phys. Rev. Lett.} \textbf{1981}, \emph{47},
  1400--1403\relax
\mciteBstWouldAddEndPuncttrue
\mciteSetBstMidEndSepPunct{\mcitedefaultmidpunct}
{\mcitedefaultendpunct}{\mcitedefaultseppunct}\relax
\EndOfBibitem
\bibitem[Ball \latin{et~al.}(1987)Ball, Weitz, Witten, and Leyvraz]{Ball1987}
Ball,~R.~C.; Weitz,~D.~A.; Witten,~T.~A.; Leyvraz,~F. Universal Kinetics in
  Reaction-Limited Aggregation. \emph{Phys. Rev. Lett.} \textbf{1987},
  \emph{58}, 274--277\relax
\mciteBstWouldAddEndPuncttrue
\mciteSetBstMidEndSepPunct{\mcitedefaultmidpunct}
{\mcitedefaultendpunct}{\mcitedefaultseppunct}\relax
\EndOfBibitem
\bibitem[Squires and Mason(2010)Squires, and Mason]{Squires2010}
Squires,~T.~M.; Mason,~T.~G. {Fluid Mechanics of Microrheology}. \emph{Annu.
  Rev. Fluid Mech.} \textbf{2010}, \emph{42}, 413--438\relax
\mciteBstWouldAddEndPuncttrue
\mciteSetBstMidEndSepPunct{\mcitedefaultmidpunct}
{\mcitedefaultendpunct}{\mcitedefaultseppunct}\relax
\EndOfBibitem
\bibitem[Cukier(1984)]{Cukier1984}
Cukier,~R.~I. Diffusion of Brownian Spheres in Semidilute Polymer Solutions.
  \emph{Macromolecules} \textbf{1984}, \emph{17}, 252--255\relax
\mciteBstWouldAddEndPuncttrue
\mciteSetBstMidEndSepPunct{\mcitedefaultmidpunct}
{\mcitedefaultendpunct}{\mcitedefaultseppunct}\relax
\EndOfBibitem
\bibitem[Phillies \latin{et~al.}(1985)Phillies, Ullmann, Ullmann, and
  Lin]{Phillies1985}
Phillies,~G. D.~J.; Ullmann,~G.~S.; Ullmann,~K.; Lin,~T. Phenomenological
  Scaling Laws for ‘‘Semidilute’’ Macromolecule Solutions from Light
  Scattering by Optical Probe Particles. \emph{J. Chem. Phys.} \textbf{1985},
  \emph{82}, 5242--5246\relax
\mciteBstWouldAddEndPuncttrue
\mciteSetBstMidEndSepPunct{\mcitedefaultmidpunct}
{\mcitedefaultendpunct}{\mcitedefaultseppunct}\relax
\EndOfBibitem
\bibitem[Cheng \latin{et~al.}(2002)Cheng, Prud'homme, and Thomas]{Cheng2002}
Cheng,~Y.; Prud'homme,~R.~K.; Thomas,~J.~L. Diffusion of Mesoscopic Probes in
  Aqueous Polymer Solutions Measured by Fluorescence Recovery after
  Photobleaching. \emph{Macromolecules} \textbf{2002}, \emph{35},
  8111--8121\relax
\mciteBstWouldAddEndPuncttrue
\mciteSetBstMidEndSepPunct{\mcitedefaultmidpunct}
{\mcitedefaultendpunct}{\mcitedefaultseppunct}\relax
\EndOfBibitem
\bibitem[Poon(1998)]{Poon1998}
Poon,~W.~C. {Phase Separation, Aggregation and Gelation in Colloid-Polymer
  Mixtures and Related Systems}. \emph{Curr. Opin. Colloid Interface Sci.}
  \textbf{1998}, \emph{3}, 593--599\relax
\mciteBstWouldAddEndPuncttrue
\mciteSetBstMidEndSepPunct{\mcitedefaultmidpunct}
{\mcitedefaultendpunct}{\mcitedefaultseppunct}\relax
\EndOfBibitem
\bibitem[Zhang \latin{et~al.}(2013)Zhang, Royall, Faers, and
  Bartlett]{Zhang2013}
Zhang,~I.; Royall,~C.~P.; Faers,~M.~A.; Bartlett,~P. {Phase Separation Dynamics
  in Colloid-Polymer Mixtures: the Effect of Interaction Range}. \emph{Soft
  Matter} \textbf{2013}, \emph{9}, 2076--2084\relax
\mciteBstWouldAddEndPuncttrue
\mciteSetBstMidEndSepPunct{\mcitedefaultmidpunct}
{\mcitedefaultendpunct}{\mcitedefaultseppunct}\relax
\EndOfBibitem
\bibitem[Carvalho \latin{et~al.}(1993)Carvalho, Tong, Huang, Witten, and
  Fetters]{Carvalho1993}
Carvalho,~B.~L.; Tong,~P.; Huang,~J.~S.; Witten,~T.~A.; Fetters,~L.~J.
  {Adsorption of End-Functionalized Polymers on Colloidal Spheres}.
  \emph{Macromolecules} \textbf{1993}, \emph{26}, 4632--4639\relax
\mciteBstWouldAddEndPuncttrue
\mciteSetBstMidEndSepPunct{\mcitedefaultmidpunct}
{\mcitedefaultendpunct}{\mcitedefaultseppunct}\relax
\EndOfBibitem
\bibitem[Zhang \latin{et~al.}(2012)Zhang, Servos, and Liu]{Zhang2012}
Zhang,~X.; Servos,~M.~R.; Liu,~J. {Ultrahigh Nanoparticle Stability Against
  Salt, pH, and Solvent with Retained Surface Accessibility via Depletion
  Stabilization}. \emph{J. Am. Chem. Soc.} \textbf{2012}, \emph{134},
  9910--9913\relax
\mciteBstWouldAddEndPuncttrue
\mciteSetBstMidEndSepPunct{\mcitedefaultmidpunct}
{\mcitedefaultendpunct}{\mcitedefaultseppunct}\relax
\EndOfBibitem
\bibitem[Fakhri \latin{et~al.}(2010)Fakhri, MacKintosh, Lounis, Cognet, and
  Pasquali]{Fakhri2010}
Fakhri,~N.; MacKintosh,~F.~C.; Lounis,~B.; Cognet,~L.; Pasquali,~M. {Brownian
  Motion of Stiff Filaments in a Crowded Environment.} \emph{Science}
  \textbf{2010}, \emph{330}, 1804--1807\relax
\mciteBstWouldAddEndPuncttrue
\mciteSetBstMidEndSepPunct{\mcitedefaultmidpunct}
{\mcitedefaultendpunct}{\mcitedefaultseppunct}\relax
\EndOfBibitem
\bibitem[Meza \latin{et~al.}(2014)Meza, Das, and Greer]{Meza2014}
Meza,~L.~R.; Das,~S.; Greer,~J.~R. {Strong, Lightweight, and Recoverable
  Three-Dimensional Ceramic Nanolattices}. \emph{Science} \textbf{2014},
  \emph{345}, 1322--1326\relax
\mciteBstWouldAddEndPuncttrue
\mciteSetBstMidEndSepPunct{\mcitedefaultmidpunct}
{\mcitedefaultendpunct}{\mcitedefaultseppunct}\relax
\EndOfBibitem
\bibitem[Cao \latin{et~al.}(2014)Cao, Sun, and Grattan]{Cao2014}
Cao,~J.; Sun,~T.; Grattan,~K. T.~V. {Gold Nanorod-Based Localized Surface
  Plasmon Resonance Biosensors: A Review}. \emph{Sensors and Actuators B:
  Chemical} \textbf{2014}, \emph{195}, 332--351\relax
\mciteBstWouldAddEndPuncttrue
\mciteSetBstMidEndSepPunct{\mcitedefaultmidpunct}
{\mcitedefaultendpunct}{\mcitedefaultseppunct}\relax
\EndOfBibitem
\end{mcitethebibliography}

\newpage
\centering
\includegraphics[height = 1.375 in]{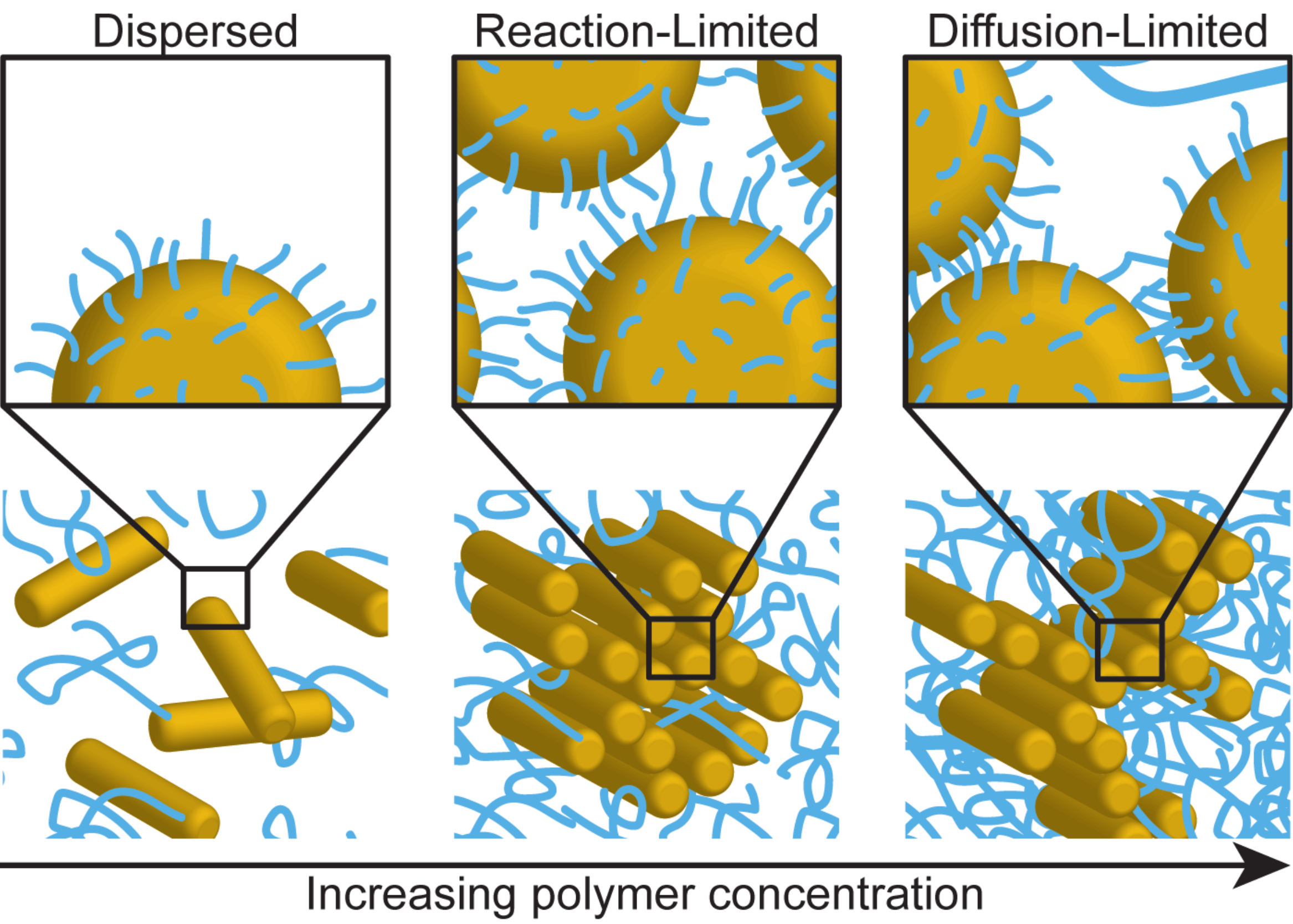}

\end{document}